\documentstyle[aps,preprint,epsf]{revtex}

\begin{document}

\tightenlines
\catcode`@=11
\def\references{%
\ifpreprintsty
\bigskip\bigskip
\hbox to\hsize{\hss\large \refname\hss}%
\else
\vskip24pt
\hrule width\hsize\relax
\vskip 1.6cm
\fi
\list{\@biblabel{\arabic{enumiv}}}%
{\labelwidth\WidestRefLabelThusFar  \labelsep4pt %
\leftmargin\labelwidth %
\advance\leftmargin\labelsep %
\ifdim\baselinestretch pt>1 pt %
\parsep  4pt\relax %
\else %
\parsep  0pt\relax %
\fi
\itemsep\parsep %
\usecounter{enumiv}%
\let\p@enumiv\@empty
\def\theenumiv{\arabic{enumiv}}%
}%
\let\newblock\relax %
\sloppy\clubpenalty4000\widowpenalty4000
\sfcode`\.=1000\relax
\ifpreprintsty\else\small\fi
}
\catcode`@=12

\hbox to \hsize{
\hfill
\vtop{\hbox{\bf MADPH-99-1145}
      \hbox{\bf Fermilab-PUB 99-341-T}
      \hbox{\bf AMES-HET 99-12}
      \hbox{hep-ph/9911524}
      \hbox{November 1999}}}

\vspace{.5in}

\begin{center}
{\large\bf Long-Baseline Study of the Leading Neutrino Oscillation\\
at a Neutrino Factory}\\[6mm]
V. Barger$^1$, S. Geer$^2$, R. Raja$^2$, and K. Whisnant$^3$\\[3mm]
\it
$^1$Department of Physics, University of Wisconsin,
Madison, WI 53706, USA\\
$^2$Fermi National Accelerator Laboratory, P.O. Box 500,
Batavia, IL 60510, USA\\
$^3$Department of Physics and Astronomy, Iowa State University,
Ames, IA 50011, USA

\end{center}

\bigskip

\begin{abstract}
Within the framework of three-flavor neutrino oscillations, we consider the
physics
potential of $\nu_e \rightarrow \nu_\mu$ appearance and
$\nu_\mu \rightarrow \nu_\mu$ survival measurements at a neutrino factory for a
leading oscillation scale $\delta m^2 \sim 3.5 \times 10^{-3}$~eV$^2$.
Event rates are evaluated versus baseline and stored muon energy, and optimal
values discussed. Over a sizeable region of oscillation parameter
space, matter effects would enable the sign of $\delta m^2$ to be determined
from a comparison of $\nu_e \rightarrow \nu_\mu$ with $\bar{\nu}_e \rightarrow
\bar{\nu}_\mu$ event rates and energy distributions. It is important,
therefore, that both positive and negative muons can be stored in the ring.
Measurements of the
$\nu_\mu \rightarrow \nu_\mu$ survival spectrum could determine the magnitude
of $\delta m^2$ and the leading oscillation amplitude with a precision of
${\cal
O}$(1\%--2\%).
\end{abstract}

\thispagestyle{empty}
\newpage

\section{Introduction}

 The SuperKamiokande (SuperK) collaboration\cite{superk} has published evidence
that muon neutrinos, produced in the earth's atmosphere by cosmic rays,
oscillate into other neutrino flavors. The $\nu_\mu$ survival probability in
vacuum for neutrinos of energy $E_\nu$~(GeV) traversing a distance $L$~(km) is
given by:
\begin{equation}
P(\nu_\mu\to\nu_\mu) = \sin^22\theta_{\rm ATM}^{\vphantom2} \sin^2(1.267\delta
m^2_{\rm ATM} L / E_\nu) \,.
\end{equation}
The SuperK results for the oscillation amplitude $\sin^22\theta_{\rm ATM}$ and
the oscillation scale $\delta m^2_{\rm ATM}$ are in accord with results
obtained from other experiments\cite{kam}.
{}From the zenith angle distribution of the muon events, which is related to
the distribution in $L/E_\nu$, the mass-squared difference scale of the
oscillations was inferred to be\cite{superk,BWW}
\begin{equation}
\delta m^2_{\rm ATM} = \left(3.5^{+3.5}_{-2.0}\right) \times 10^{-3}\rm\, eV^2
\label{eq:dm2}
\end{equation}
and the amplitude of the oscillations was found to be maximal or nearly
maximal,
\begin{equation}
\sin^2 2\theta_{\rm ATM} = 1^{+0}_{-0.2}  \,.
\end{equation}
No zenith angle dependence was observed by SuperK for electron events, so it is
concluded that electron-neutrinos do not undergo appreciable oscillations at
the $\delta m^2_{\rm ATM}$ scale of Eq.~(\ref{eq:dm2}) and that the
muon-neutrinos
oscillate dominantly to some neutrino flavor other than electron-neutrinos.
This interpretation is consistent with the stringent lower limits on the
electron-neutrino survival probability $P(\bar\nu_e\to\bar\nu_e)$ from the
CHOOZ reactor neutrino experiment\cite{chooz}.
Thus it is inferred that atmospheric
muon-neutrinos oscillate to tau-neutrinos or to a new sterile neutrino species
that has no Standard Model interactions. In the latter case, matter effects
would be expected to
distort the zenith angle distributions at large zenith angles. The SuperK
data presently disfavor muon-neutrino oscillations to sterile neutrinos by
two standard deviations\cite{superk}.

For $\delta m^2$ given by Eq.~(\ref{eq:dm2}), the first minimum in the
survival probability occurs at
\begin{equation}
L/E_\nu \simeq 350^{+480}_{-170}\rm\ km/GeV \,.  \label{eq:minprob}
\end{equation}
The detector averaged $ P(\nu_\mu\to\nu_\mu) $ measurement by
SuperK does not resolve
this minimum because of the smearing over $L$ and inferred $E_\nu$ values.
Accelerator-based experiments are thus essential to establish the existence of
the minimum in the survival probability.
The K2K\cite{k2k} experiment from KEK to SuperK,
already underway, has $L=250$~km and average neutrino energy $\left< E_\nu
\right> = 1.4$~GeV, with an $L/E_\nu$ range of 125--250~km/GeV.
The MINOS\cite{minos} experiment from
Fermilab to Soudan has a baseline $L = 732$~km and 3 beam options with $\left<
E_\nu \right> = 3$, 6 and 12~GeV, giving $L/E_\nu = 50$--250~km/GeV. Other
experiments from CERN to detectors in Gran Sasso with a baseline $L = 743$~km
are in the final design stages.
The energy dependence
of the charged-current rate, the neutral-current rate, and $\tau$-appearance
with higher energy beams will test the neutrino oscillation hypothesis. The
value of
$\delta m^2$ will be measured in these long-baseline experiments to higher
precision than possible with atmospheric neutrinos. For example, MINOS may
ultimately be able~\cite{thesis}
to make a 10\% determination of $\delta m^2$.

Thus the next generation of accelerator long-baseline neutrino experiments
   are expected to firmly establish neutrino oscillations and improve our
knowledge of $\delta m^2$. However, the dominant flux components in these
accelerator experiments are $\nu_\mu$ and $\bar{\nu}_\mu$. To make   further
progress in determining all of the parameters describing the   oscillations it
is desirable to also have $\nu_e$ and $\bar{\nu}_e$   beams, in addition to
beams of higher intensity. With this in mind, it   has been pointed
out~\cite{geerconf,geer} that if an intense muon source of   the type   being
developed for a possible future muon collider~\cite{mucoll} is used   together
with a muon storage ring having long straight sections, the   resulting
``neutrino factory" would produce intense beams containing   $\nu_e$
($\bar{\nu}_e$) as well as $\bar{\nu}_\mu$ ($\nu_\mu$).   In the original
neutrino factory proposal~\cite{geer} it was shown that   if the storage ring
was tilted down at a large angle, the neutrino beams   would be sufficiently
intense to produce thousands of interactions in a   reasonable sized detector
on the other side of the Earth. Neutrino   factories have caught the attention
of the community, and several groups   are developing the concept with BNL,
CERN, FNAL, and KEK all being   considered as possible
sites~\cite{derujula,camp,nu-factory}.

In Ref.~\cite{BGW} we explored long-baseline neutrino physics at a neutrino
factory within the framework of two-flavor oscillations.   In the present paper
we extend our analysis of the physics potential at neutrino factories,
presenting results within the framework of three-flavor oscillations. In
particular we discuss: (i)~the precise determination of 
$\delta m_{\rm ATM}^2$ ($\sim 1\%-2\%$) from the measured 
$\nu_\mu\to\nu_\mu$ survival probability;
(ii)~a proof that the sign of $\delta m_{\rm ATM}^2$ can in principle
be extracted from $\nu_e\to\nu_\mu$ and
$\nu_\mu\to \nu_e$ appearance measurements by exploiting matter effects
that modify these
oscillation probabilities; and (iii)~the optimal $L/E$ and methodology for
the measurements.
Table~\ref{table:baselines} summarizes the baselines,
average electron density in the earth, and
the dip angle for various possibilities. Our present analysis focuses on
long-baseline experiments with $L = 732$~km (Fermilab $\rightarrow$ Soudan),
$L = 2800$~km and 2900~km (Note: Fermilab $\rightarrow$ SLAC $\sim$2900~km,
Fermilab $\rightarrow$ Seattle, Washington $\sim$2700~km),
and $L = 7332$~km (Fermilab $\rightarrow$ Gran~Sasso).

\section{Leading Neutrino Oscillation}
\label{sec:leading-osc}

There are experimental indications of neutrino oscillation effects from
the LSND accelerator experiment\cite{lsnd}, from the atmospheric
neutrino anomaly\cite{superk,kam,old}, and from the solar neutrino
deficit\cite{solar1,solar2}. Three neutrino mass-squared differences are
required to completely explain all these phenomena. However, three neutrinos
 provide only two distinct $\delta m^2$ scales. Therefore,  a sterile neutrino
must be invoked if all the experimental indications are real. Since the
significance of the LSND effect is not at the discovery level, a common
approach is to set this anomaly aside until it is confirmed or rejected
by the forthcoming Fermilab Mini-BooNE experiment\cite{miniboone} and analyze
the solar and  atmospheric neutrino deficits within the framework of
three-neutrino oscillations. This is the route followed
in the present analysis. Thus we take the atmospheric $\delta m^2$ sale to be
the leading oscillation.

With three neutrinos, the flavor eigenstates $\nu_\alpha\ (\alpha=e, \mu,
\tau)$ are related to the mass eigenstates $\nu_j\ (j=1,2,3)$ in vacuum by
\begin{equation}
\nu_\alpha = \sum_j U_{\alpha j} \nu_j \,,
\end{equation}
where $U$ is the unitary Maki-Nakagawa-Sakata (MNS) mixing matrix\cite{mns}. We
parametrize $U$ by
\begin{equation}
U
= \left( \begin{array}{ccc}
  c_{13} c_{12}       & c_{13} s_{12}  & s_{13} e^{-i\delta} \\
- c_{23} s_{12} - s_{13} s_{23} c_{12} e^{i\delta}
& c_{23} c_{12} - s_{13} s_{23} s_{12} e^{i\delta}
& c_{13} s_{23} \\
    s_{23} s_{12} - s_{13} c_{23} c_{12} e^{i\delta}
& - s_{23} c_{12} - s_{13} c_{23} s_{12} e^{i\delta}
& c_{13} c_{23} \\
\end{array} \right) \,,
\end{equation}
where $c_{jk} \equiv \cos\theta_{jk}$ and $s_{jk} \equiv \sin\theta_{jk}$.
For Majorana neutrinos, $U$ contains two further multiplicative phase
factors that modify diagonal entries of $U$, but these do not enter in
oscillation phenomena.

The importance of forward scattering of neutrinos on electrons in the
propagation of neutrinos through matter was first pointed out in
Ref.~\cite{wolf}. The existence of resonance effects in propagation of
neutrinos through the earth in a constant density approximation was then
discovered in Ref.~\cite{bppw}. Corrections to the magnitude and sign of the
coherent amplitude given by Ref.~\cite{wolf} were made in
Refs.~\cite{bppw,langacker}. The
effects of matter are prominent in neutrino oscillation
solutions\cite{miksmirnov} to
the solar neutrino anomaly\cite{solar1,solar2}. Early calculations for long-baseline experiments
were presented in Ref.~\cite{burstein}. Many long-baseline analytical and
numerical studies have subsequently been made that include various
refinements~\cite{lipari,petcov,zaglauer,fuller,pantaleone,matter,%
ohlsson}.

The propagation of three neutrinos through matter is described by the evolution
equation
\begin{equation}
i{d\nu_\alpha\over dx} = \sum_\beta \left[ \left( \sum_j U_{\alpha j} U_{\beta
j}^* {m_j^2\over 2E_\nu} \right) + {A\over 2E_\nu} \delta_{\alpha e}
\delta_{\beta e} \right] \nu_\beta \,,  \label{eq:prop}
\end{equation}
where $x=ct$ and $A/(2E_\nu)$ is the amplitude for
coherent forward charged-current scattering of $\nu_e$ on electrons,
\begin{equation}
A = 2\sqrt2 G_F Y_e\rho E_\nu = 1.52 \times 10^{-4}{\rm\,eV^2} Y_e
\rho({\rm\,g/cm^3}) E({\rm\,GeV}) \,.
\label{eq:A}
\end{equation}
Here $Y_e(x)$ is the electron fraction and $\rho(x)$ is the matter density. For
neutrino trajectories that pass through the earth's crust, the average density
is typically of
order 3~gm/cm$^3$ and $Y_e \simeq 0.5$. The propagation Eq.~(\ref{eq:prop})
can be re-expressed in terms of mass-squared differences as
\begin{equation}
i{d\nu_\alpha\over dx} = \sum_\beta {1\over2E_\nu} \left[
  \delta m_{31}^2 U_{\alpha 3} U_{\beta 3}^*
+ \delta m_{21}^2 U_{\alpha 2} U_{\beta 2}^*
+ A \delta_{\alpha e} \delta_{\beta e} \right]
\nu_\beta\,.  \label{eq:prop2}
\end{equation}
This evolution equation can be solved numerically for given input values of the
$\delta m^2$ and mixing matrix elements.

In our analysis we obtain numerical solutions of Eq.~(\ref{eq:prop2}) for
three-neutrino oscillations taking into account the dependence of the density
on depth using the Preliminary Reference Earth Model~\cite{PREM}. However,
it is instructive to first consider an approximate analytic
solution\cite{pantaleone} based on the $\delta m^2$ hierarchy
\begin{equation}
| \delta m_{32} |^2 \simeq | \delta m_{31} |^2 \gg | \delta m_{21} |^2 \,.
\end{equation}
implied by oscillation solutions to the atmospheric and solar neutrino
anomalies. Then, for short enough distances, the $\delta m_{21}^2$ term in
Eq.~(\ref{eq:prop2}) can be dropped and the other two $\delta m^2$ set equal,
\begin{equation}
\delta m_{32}^2 = \delta m_{31}^2 \equiv \delta m^2 \,.
\end{equation}
The evolution equations are then
\begin{equation}
i {d\over dt}
\left( \begin{array}{c} \nu_e \\ \nu_\mu \\ \nu_\tau  \end{array} \right)
= {\delta m^2\over 2E}
\left( \begin{array}{ccc}
{A\over \delta m^2} + |U_{e3}|^2 & U_{e3}U_{\mu3}^* & U_{e3}U_{\tau3}^* \\
U_{e3}^*U_{\mu3} & |U_{\mu3}|^2 & U_{\mu3}U_{\tau3}^* \\
U_{e3}^*U_{\tau3} & U_{\mu3}^*U_{\tau3} & |U_{\tau3}|^2
\end{array} \right)
\left( \begin{array}{c} \nu_e \\ \nu_\mu \\ \nu_\tau \end{array} \right)
\,.
\end{equation}
The propagation matrix has eigenvalues
\begin{equation}
\lambda_1 = 0 \,, \qquad
\lambda_{2} =
{\delta m^2\over 4E}\left[ 1 + {A\over\delta m^2} - S \right] \,, \qquad
\lambda_{3} =
{\delta m^2\over 4E}\left[ 1 + {A\over\delta m^2} + S \right] \,,
\label{eq:lambda}
\end{equation}
where
\begin{equation}
S \equiv \sqrt{ \left( {A\over\delta m^2}-\cos2\theta_{13} \right)^2
+ \sin^2 2\theta_{13}} \,.
\label{eq:S}
\end{equation}
For propagation through matter of constant density, the flavor eigenstates are
related to the mass eigenstates $\nu_j^m$ by
\begin{equation}
\nu_\alpha = \sum U_{\alpha j}^m \nu_j^m \,,
\end{equation}
where
\begin{equation}
U^m = \left( \begin{array}{ccc}
      0 &  c_{13}^m & s_{13}^m \\
-c_{23} & -s_{13}^m s_{23} & c_{13}^m s_{23} \\
 s_{23} & -s_{13}^m c_{23} & c_{13}^m c_{23}
\end{array} \right)
\end{equation}
and $\theta_{13}^m$ is related to $\theta_{13}$ by
\begin{equation}
\tan 2\theta_{13}^m = {\sin^2 2\theta_{13}\over \cos2\theta_{13}
- {A\over \delta m^2}} \,. \label{eq:tan}
\end{equation}
We note that $U^m$ has the form of the vacuum $U$ with the substitutions
\begin{equation}
\theta_{13}\to\theta_{13}^m\,,  \quad
\theta_{23}\to\theta_{23}\,,\quad 
\theta_{12}\to\pi/2\,,\quad
\delta = 0 \,.
\end{equation}
There are only two mixing angles because the oscillations associated with the
solar scale ($\delta m_{21}^2$) have not yet developed. The effective phase
angle $\delta$ vanishes, so CP violation is not possible in the leading
oscillation, even after matter effects are included\cite{bpw}.
Equation~(\ref{eq:tan})
implies that
\begin{equation}
\sin^2 2\theta_{13}^m = {\sin^2 2\theta_{13}\over
\left({A\over\delta m^2} - \cos 2\theta_{13} \right)^2
+ \sin^2 2\theta_{13}} \,. \label{eq:sin}
\end{equation}
Thus there is a resonant enhancement for
\begin{equation}
A = \delta m^2 \cos2\theta_{13}
\end{equation}
or equivalently
\begin{equation}
E_\nu \approx 15{\rm\ GeV} \left(\delta m^2 \over 3.5\times
10^{-3}{\rm\,eV^2}\right) \left( 1.5{\rm\ g/cm^3}\over \rho Y_e \right)
\cos2\theta_{13} \,. \label{eq:Enu}
\end{equation}
The resonance occurs only for positive $\delta m^2$. For negative
$\delta m^2$ the oscillation amplitude in (\ref{eq:sin}) is smaller than the
vacuum oscillation amplitude. Thus the matter effects give us a way in
principle to determine the sign of $\delta m^2$.

For a constant Earth density profile, the transition probabilities for a given
neutrino flavor are given by
\begin{equation}
P(\nu_\alpha\to \nu_\beta) =
- 4 \sum_{j<k} U_{\alpha j}^m U_{\alpha k}^m U_{\beta j}^m U_{\beta k}^m
\sin^2\Delta_{kj}^m \,,
\end{equation}
where
\begin{equation}
\Delta_{kj}^m=L(\lambda_k-\lambda_j)/2
\end{equation}
and the $\lambda_j$ are the eigenvalues of the neutrino matrix in
Eq.~(\ref{eq:lambda}). The transition probabilities in the leading oscillation
approximation are
\begin{eqnarray}
P(\nu_e\to \nu_\mu) &=& s_{23}^2 \sin^2 2\theta_{13}^m \sin^2\Delta_{32}^m \,,
\nonumber\\
P(\nu_e\to \nu_\tau) &=& c_{23}^2 \sin^2 2\theta_{13}^m \sin^2\Delta_{32}^m \,,
\label{eq:probs}\\
P(\nu_\mu\to \nu_\tau) &=& \sin^2 2\theta_{23} \left[
 (\cos\theta_{13}^m)^2 \sin^2\Delta_{21}^m
+(\sin\theta_{13}^m)^2 \sin^2\Delta_{31}^m
-(\sin\theta_{13}^m\cos\theta_{13}^m)^2 \sin^2\Delta_{32}^m \right]
\,. \nonumber
\end{eqnarray}
Here the oscillation arguments are
\begin{equation}
\Delta_{32}^m = \Delta_0 S \,,\qquad
\Delta_{31}^m = \Delta_0 {1\over2} \left[ 1+{A\over\delta m^2}+S \right]
\,, \qquad
\Delta_{21}^m = \Delta_0 {1\over2} \left[ 1+{A\over\delta m^2}-S \right]
\,, \label{eq:arg}
\end{equation}
where $S$ is given by Eq.~(\ref{eq:S}) and
\begin{equation}
\Delta_0 = {\delta m^2_{32} L\over 4E} = 1.267 {\delta m^2_{32}
{\rm\,(eV^2)} \; L {\rm\ (km)} \over E_\nu {\rm\ (GeV)}}
\,.
\label{eq:arg0}
\end{equation}

The $\Delta_{21}^m$ term in $P(\nu_\mu\to\nu_\tau)$ must be retained here
because it is not necessarily negligible compared to $\Delta_{31}^m$, due to
matter effects. From Eqs.~(\ref{eq:sin}), (\ref{eq:probs}) and (\ref{eq:arg}),
we see in the limit $\theta_{13}\to0$ that $\theta_{13}^m\to0$ and the $\nu_e$
transition probabilities vanish. We see from Eq.~(\ref{eq:probs}) that matter
effects are possible for $\nu_\mu\to\nu_\tau$ oscillations in a three neutrino
model. However, these matter effects disappear in the $\theta_{13}=0$ limit,
where
\begin{equation}
S = \left| 1 - {A\over\delta m^2_{\rm ATM}} \right| \,,
\label{eq:approxS}
\end{equation}
and
\begin{equation}
P(\nu_\mu\to \nu_\tau) = \sin^2 2\theta_{23} \sin^2(\Delta_0) \,.
\label{eq:nomatter}
\end{equation}

To have substantial $\tau$ event rates from $\nu_\mu\to\nu_\tau$ oscillations
requires high energy neutrino beams because of the kinematic suppression of the
$\tau$ production cross section near threshold. The $\nu_e\to\nu_\mu$ and
$\nu_\mu\to\nu_\mu$ oscillation probabilities can be studied with either low or
high energy beams and we focus our attention on them in the rest of this paper.
A non-zero value of $\theta_{13}$ is essential for the occurrence of these
oscillations and for the observation of matter effects.

\section{Neutrino Fluxes and Cross-Sections}

The distribution $n(x,\Omega) = d^2N/dxd\Omega$ of neutrinos from the decays of
an ensemble of polarized negatively-charged muons in the muon rest frame is given by
\begin{eqnarray}
n_{\nu_\mu}(x,\Omega) &=& {2x^2\over4\pi}
 \left[ (3-2x) + (1-2x) P_\mu \cos\theta \right] \,,
\label{eq:n_numu}\\
n_{\bar\nu_e}(x,\Omega) &=&  {12x^2\over4\pi}
\left[ (1-x) + (1-x) P_\mu\cos\theta \right] \,,
\label{eq:n_nue}
\end{eqnarray}
where $x\equiv 2E_\nu/m_\mu$, $\theta$ is the angle between the neutrino
momentum vector and the muon spin direction and $P_\mu$ is the average muon
polarization along the beam direction. The corresponding distributions for
$\bar\nu_\mu$ and $\nu_e$ from $\mu^+\to e^+\bar\nu_\mu \nu_e$ are obtained by
the replacement $P_{\mu^-} \to -P_{\mu^+}$ in Eqs.~(\ref{eq:n_numu}) and
(\ref{eq:n_nue}). Only neutrinos emitted in the extreme forward
direction ($\cos\theta\simeq1$) are relevant to the neutrino flux for
long-baseline experiments; in this limit
\begin{equation}
E_\nu = x E_\mu 
\end{equation}
in the lab frame.
The flux at a distance $L$ from the storage ring can be approximated by
\begin{equation}
\Phi \simeq {n_0\gamma^2\over \pi L^2} \,,
\end{equation}
where $\gamma = E_\mu/m_\mu$ and $n_0$ is the number of neutrinos (or
antineutrinos) in a given beam. The charged-current (CC) interaction
cross-sections in the detector
grow approximately linearly with the neutrino energy
\begin{eqnarray}
\sigma_{\nu N} &\simeq& 0.67\times 10^{-38}{\rm\,cm^2} \times E_\nu {\rm\
(GeV)}\,. \\
\sigma_{\bar\nu N} &\simeq& 0.34\times 10^{-38}{\rm\,cm^2} \times E_{\bar\nu}
\rm\ (GeV) \,. \label{eq:sigmanubar}
\end{eqnarray}
Thus, the event rates in the absence of oscillations are
\begin{equation}
N \sim (E_\mu)^3 /L^2 \,.
\label{eq:rate}
\end{equation}
The $\nu_\mu$ ($\bar\nu_\mu$) event energies peak at the stored $\mu^-$
($\mu^+$) energy while the $\bar\nu_e$ ($\nu_e$) energies peak at about
3/4 of the stored $\mu^-$ ($\mu^+$) energy. The lepton spectra from the
charged-current interactions can be obtained by folding the event rates with
the $d\sigma/dy$ distributions, where $y=1-E_\ell/E_\nu$.

For no oscillations, the average observed neutrino energies would be
\begin{eqnarray}
\left< E_{\nu_\mu} \right> &=& 0.7 E_{\mu^-} \,,\\
\left< E_{\bar\nu_e} \right> &=& 0.6 E_{\mu^-} \,.
\end{eqnarray}
Thus, from Eq.~(\ref{eq:minprob}), the first maximum in the leading oscillation
occurs for
\begin{equation}
L/E_\mu \sim 230^{+310}_{-110}\rm\ km/GeV \,. \label{eq:L}
\end{equation}
For maximal oscillation effects the baseline and stored energy should be chosen
to satisfy Eq.~(\ref{eq:L}). In order to have sufficient event rates at
$>1000$~km distances, the lowest practical muon energy is about 10~GeV. The
corresponding $L$ from Eq.~(\ref{eq:L}) is close to the 2900~km baseline from
Fermilab to SLAC.

\section{Oscillation Channels}

The decays of stored muons along the straight sections of the storage ring give
well-collimated neutrino beams of specified flavor composition. For stored
$\mu^-$, the $\mu^-\to\nu_\mu \bar\nu_e e^-$ decays give $\nu_\mu$ and
$\bar\nu_e$ beams with the known energy distributions given by
Eq.~(\ref{eq:n_numu}) (see Fig.~4 of \cite{BGW}). In this paper we restrict our
attention to detection of muons. The relevant oscillation signatures for
stored $\mu^-$ are given in Table~\ref{tab:osc-probs}. Corresponding signatures
for stored $\mu^+$ are obtained by the interchange $\nu\leftrightarrow \bar\nu$
and changing the signs of the electric charges of the leptons.

The oscillations to $\nu_\tau$ feed the same final state as oscillations to
$\nu_\mu$ through the decays $\tau\to\mu$ of the produced $\tau$-leptons. Thus
detection of muons determines linear combinations of oscillation probabilities
\begin{eqnarray}
N (\mu^-) &=& \left< \Phi P(\nu_\mu\to\nu_\mu)
\sigma(\nu_\mu\to\mu^-) \right> + \left< \Phi P(\nu_\mu\to\nu_\tau)
\sigma(\nu_\tau\to\tau^-) \right> {\rm BF}(\tau^-\to\mu^-) \,,\\
N (\mu^+) &=& \left< \Phi P(\bar\nu_e\to\bar\nu_\mu)
\sigma(\bar\nu_\mu\to\mu^+) \right> + \left< \Phi P(\bar\nu_e\to\bar\nu_\tau)
\sigma(\bar\nu_\tau\to\tau^+) \right> {\rm BF}(\tau^+\to\mu^+) \,,
\end{eqnarray}
where $\Phi$ is the flux and $\left<\phantom{P}\right>$ denotes a spectrum
average and
integration over final state energies. Neglecting the small matter
effects in $\nu_\mu\to\nu_\tau$, the leading oscillation rates from
Eqs.~(\ref{eq:probs}) and (\ref{eq:nomatter}) would be
\begin{eqnarray}
N (\mu^-) &=& \left< \Phi \left[ 1
- s_{23}^2 \sin^2 2\theta_{13}^m \sin^2\Delta_{32}^m
- \sin^2 2\theta_{23} \sin^2\Delta_0 \right]
\; \sigma(\nu_\mu\to\mu^-) \right> \nonumber\\
&& {} + \sin^2 2\theta_{23} \left< \Phi \sin^2\Delta_{32} \;
\sigma(\nu_\tau\to\tau^-) \right> {\rm BF}(\tau^-\to\mu^-) \,,\\
N (\mu^+) &=& s_{23}^2 \sin^2 2\theta_{13}^m
\left< \Phi \sin^2 \Delta_{32}^m \; \sigma(\bar\nu_\mu\to\mu^+) \right>
\nonumber\\
&& {} + c_{23}^2\sin^2 2\theta_{13}^m \left< \Phi \sin^2 \Delta_{32}^m \;
\sigma(\bar\nu_\tau\to\tau^+) \right> {\rm BF} (\tau^+\to\mu^+) \,.
\end{eqnarray}
The $\tau$-contributions will be suppressed at low stored muon energies
(${\sim}10$~GeV) by the kinematic suppression of the tau cross section near the
threshold for tau production.
The tau neutrino interactions can be identified by 
(i)~kinematics ($p_T$ relative
to the neutrino beam direction and the $y$-distribution of the decay muon), or 
(ii) direct evidence for $\tau\to 1$-prong or $\tau\to 3$-prong decays 
by detecting kinks or displaced vertices in, for example, emulsion detectors, 
or by imaging Cherenkov rings that the tau generates in
C$_6$\,F$_{14}$ liquid~\cite{forty}.
Henceforth we concentrate our analysis strictly on primary muons from $\nu_\mu$
under the assumption that the contributions of muons from tau decay can be
resolved.

\section{Predicted Neutrino Oscillation Event Rates}

For a quantitative analysis of neutrino oscillation event rates in
long-baseline experiments, we begin with the following set of oscillation
parameters:
\begin{eqnarray}
|\delta m_{32}^2 | &=& 3.5\times 10^{-3}\rm\,eV^2\nonumber \\
|\delta m_{21}^2 | &=& 5\times 10^{-5}\rm\,eV^2\nonumber \\
s_{13} &=& 0.10 \qquad (\sin^2 2\theta_{13} = 0.04) \nonumber\\
s_{23} &=& 0.71 \qquad (\sin^2 2\theta_{23} = 1.00)\label{eq:params} \\
s_{12} &=& 0.53 \qquad (\sin^2 2\theta_{12} = 0.80) \nonumber\\
\delta &=& 0 \nonumber
\end{eqnarray}
The values of $\delta m_{32}^2$
     and $s_{23}$ govern the atmospheric neutrino oscillations. The values of
     $\delta m_{21}^2$ and $s_{12}$ control the solar neutrino oscillations.
 We select the large-angle matter solution\cite{BKSV-etal} because it is the
most optimistic choice for obtaining any effects in long-baseline experiments
     from the subleading $\delta m_{21}^2$ scale. The value of $s_{13}$
determines the
size of the matter effects for the leading oscillations. The value of $\delta
m_{31}^2$ is determined from $\delta m_{32}^2$ and $\delta m_{21}^2$ by the sum
rule
\begin{equation}
\delta m_{31}^2 = \delta m_{32}^2 + \delta m_{21}^2 \,.
\end{equation}
The subleading oscillation effects associated with the $\delta
m_{21}^2$ in Eq.~(\ref{eq:params}) are usually small in comparison with the
leading oscillation, so either of the other two solar solution possibilities
(small angle matter or large-angle vacuum oscillations) would lead to
results similar to the representative case above; we will remark below when
the subleading oscillation effects are appreciable. The
long-baseline oscillations are controlled by the three parameters $\delta
m_{32}^2,\ s_{23}$ and $s_{13}$. The matter effects in the leading
oscillation are critically dependent on $s_{13}$.

\begin{figure}

\centering\leavevmode
\epsfxsize=6in\epsffile{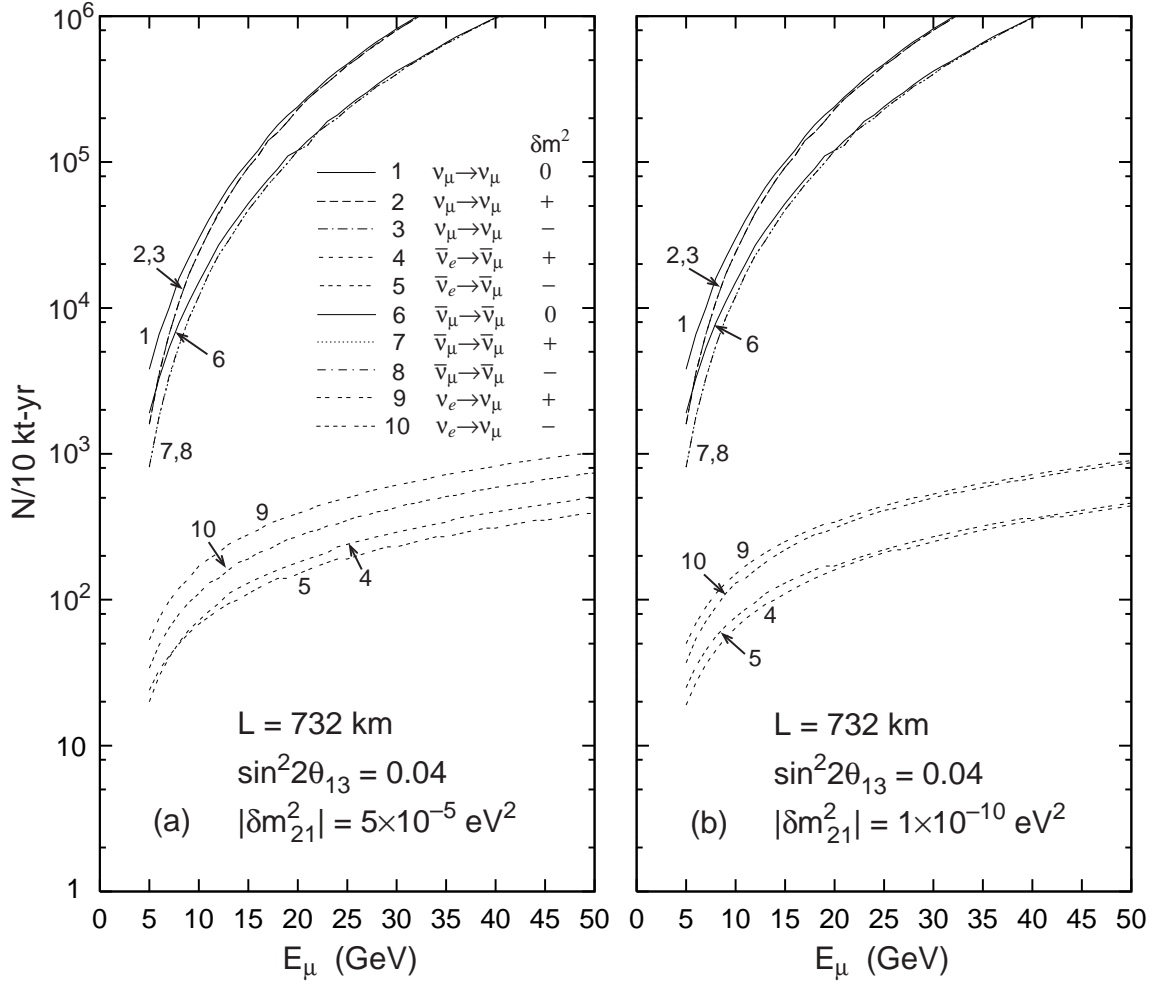}
\medskip
\caption[]{ \label{fig:732-vs-E} Muon survival and appearance events per
10~kt-yr versus muon energy at $L=732$~km, assuming $2\times10^{20}$
muon decays per year. The oscillation parameters are those given in
Eq.~(\ref{eq:params}) and (a) $\delta m^2_{21} =
5\times10^{-5}$~eV$^2$ and (b) $\delta m^2_{21} \sim 1\times 10^{-10}$~eV$^2$.
The upper (lower) solid curve shows the rates without oscillations for
neutrinos (antineutrinos). For the rates with oscillations, the results
are shown for both signs of $\delta m^2_{32}$.}
\end{figure}

We calculate neutrino event rates, with and without oscillations, for a
neutrino factory with $2\times10^{20}$ stored muons/year and a
10~kiloton detector.
For the neutrino flux, energy distributions and cross sections we use
Eqs.~(\ref{eq:n_numu})--(\ref{eq:sigmanubar}). The oscillation
probabilities are calculated by integrating Eq.~(\ref{eq:prop}) numerically
 from the source to the detector using a Runge--Kutta
method, and averaging over the neutrino energy distribution. For the
Earth's density we use the results of the Preliminary Reference Earth
Model~\cite{PREM}.
 In these initial exploratory results we neglect
the stored muon beam momentum-spread and angular-divergence, which we later
incorporate in Sec.~\ref{sec:detail}. Fig.~\ref{fig:732-vs-E}a
shows the event rates at $L=732$~km versus stored muon (or antimuon)
energy $E_\mu$ for the oscillation parameters in
Eq.~(\ref{eq:params}).
We show in
Fig.~\ref{fig:732-vs-E}b the corresponding event rates
with $\delta m^2_{21} \sim
10^{-10}$~eV$^2$ for the subleading oscillation scale, i.e., appropriate for
vacuum long-wavelength solar
oscillations. We note the following:

\begin{enumerate}

\item[(i)] For the appearance channels $\nu_e \to \nu_\mu$ and $\bar\nu_e
\to \bar\nu_\mu$, the main difference in rates comes from the difference
in the neutrino and antineutrino cross sections. Matter effects, which
arise when the sign of $\delta m^2_{32}$ is changed, are seen to be
relatively small in Fig.~\ref{fig:732-vs-E}b.  This is to be expected
since the characteristic wavelength for matter effects at these energies is of
order the Earth's diameter, which is much larger than 732~km.

\item[(ii)] The effects of the subleading mass scale, $\delta m^2_{21}$,
are evident by comparing the appearance channels in
Fig.~\ref{fig:732-vs-E}a to those in Fig.~\ref{fig:732-vs-E}b. For the
large-angle
MSW solution, $\delta m^2_{21} = 5\times10^{-5}$~eV$^2$, the interplay
of the subleading scale with matter and the sign of $\delta m^2_{32}$
affects the rates at the 20\% level. We have found that this effect
increases as $\sin^2 2\theta_{13} \rightarrow 0$, when the subleading
leading oscillation, with its larger oscillation amplitude, can begin to
compete with the leading oscillation, which has a small amplitude in
this limit\cite{peresmirnov}.

\item[(iii)] At lower muon energies (10~GeV or less) the survival
probabilities for the $\nu_\mu \to \nu_\mu$ and
$\bar\nu_\mu \to \bar\nu_\mu$ channels decrease with decreasing energy,
but do not reach their minimum values, on average, for $E_\mu > 5$~GeV.

\item[(iv)] The survival rates do not depend appreciably on the sign
of $\delta m^2_{32}$, which means that matter effects are small in these
channels at this distance, as is the case for the appearance
channels.

\end{enumerate}

\begin{figure}

\centering\leavevmode
\epsfxsize=6in\epsffile{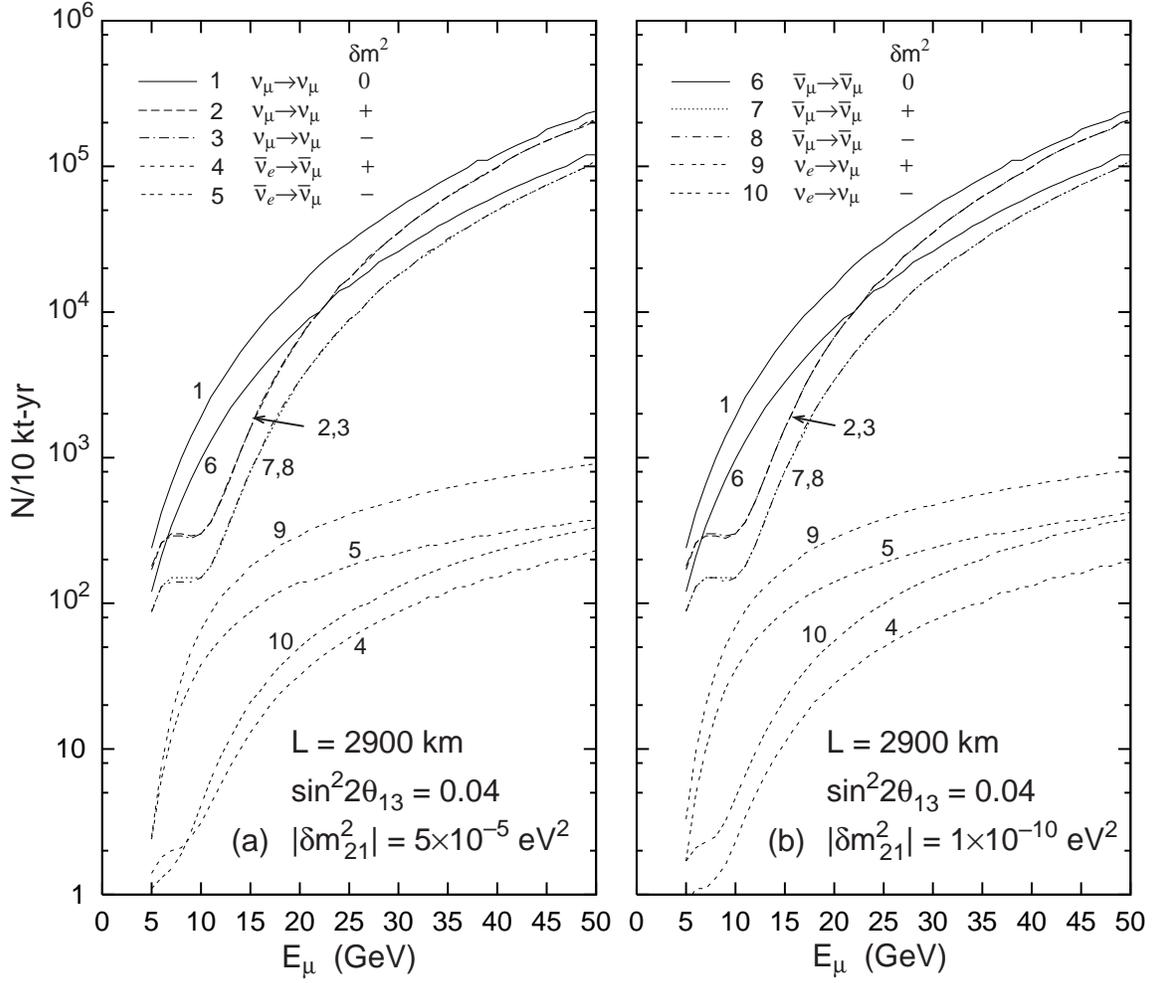}
\caption[]{ \label{fig:2900-vs-E} Same as Fig.~\ref{fig:732-vs-E} except
$L=2900$~km.}
\end{figure}

In Fig.~\ref{fig:2900-vs-E} we show similar results for $L=2900$~km, from
which we conclude:

\begin{enumerate}

\item[(i)] The largest
     event rate suppression in the survival channels occurs at energies $E_\mu
\sim 10$~GeV.

\item[(ii)] Changing the sign of $\delta m_{32}^2$ causes noticeable
changes in the event rates in the appearance channels $\nu_e\to\nu_\mu$
and $\bar\nu_e\to\bar\nu_\mu$. By comparing Fig.~\ref{fig:2900-vs-E}a with
Fig.~\ref{fig:2900-vs-E}b we see that the subleading oscillation scale has a
smaller relative effect for $L=2900$~km than for $L=732$~km
(Fig.~\ref{fig:732-vs-E}), so these changes must largely arise
from matter effects.

\end{enumerate}

\begin{figure}

\centering\leavevmode
\epsfxsize=6in\epsffile{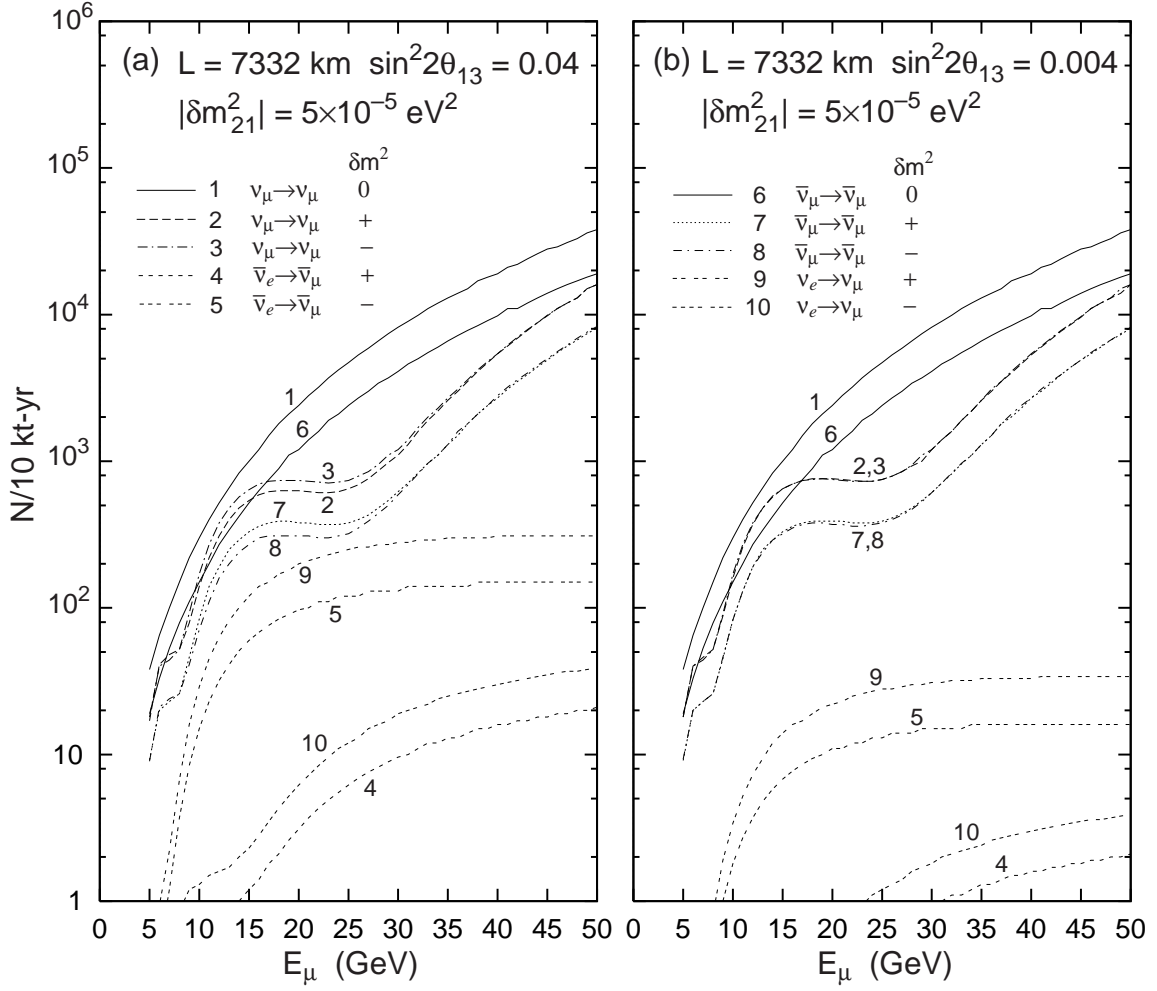}
\medskip
\caption[]{ \label{fig:7332-vs-E} Same as Fig.~\ref{fig:732-vs-E} except
$L=7332$~km, $\delta m^2_{21}=5\times10^{-5}$~eV$^2$, and
(a) $\sin^22\theta_{13}=0.04$ and (b) $\sin^22\theta_{13}=0.004$.}
\end{figure}

In Fig.~\ref{fig:7332-vs-E} we show similar results for $L=7332$~km, from
which we conclude:

\begin{enumerate}

\item[(i)] Matter effects have now grown quite large, as evidenced by
the large variation in the appearance event rates in
Fig.~\ref{fig:7332-vs-E}a when the sign of $\delta m^2_{32}$ is changed. We
have found that the results using the large-angle MSW and vacuum solar scales
for the subleading oscillations are nearly identical at this distance.

\item[(ii)] A minimum in the survival probability is
clearly visible near $E_\mu=25$~GeV. Also, the {\it survival} rates
are now sensitive to the sign of $\delta m^2_{32}$ for $E_\mu \sim
15-25$~GeV, indicating matter effects. Unfortunately, the number of
events in one of the appearance channels (either $\nu_e \rightarrow
\nu_\mu$ or $\bar\nu_e \rightarrow \bar\nu_\mu$, depending on the sign
of $\delta m^2_{32}$) falls below 10 for these energies, even for
$\sin^2 2\theta_{13} = 0.04$. In general, the event rates at $7332$~km
are below those at $2900$~km due to a lower flux.

\item[(iii)] By comparing Fig.~\ref{fig:7332-vs-E}a with
Fig.~\ref{fig:7332-vs-E}b we see that matter effects in the
survival channels are drastically reduced as $\sin^2 2\theta_{13}$
is decreased, as predicted in Eq.~(\ref{eq:nomatter}).

\item[(iv)] The number of events in the appearance channels is
sufficiently large to allow
a determination
of the sign of $\delta m^2_{32}$, at least for the $L=2900$~km and 7332~km
distances.

\end{enumerate}

\begin{figure}

\centering\leavevmode
\epsfxsize=6in\epsffile{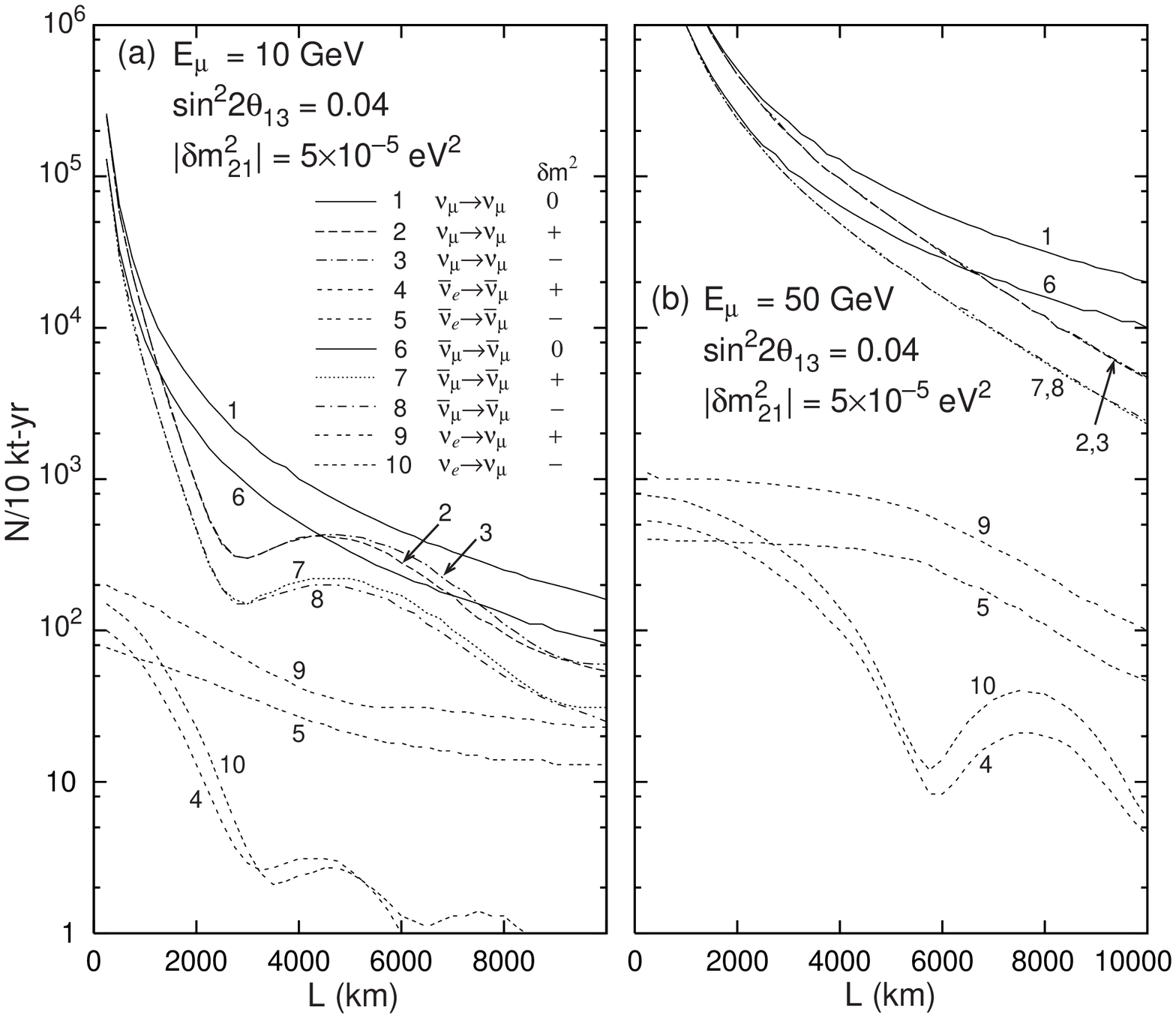}
\medskip
\caption[]{ \label{fig:fixedE-vs-L} Muon survival and appearance rates
per 10~kt-yr versus detector distance for (a) $E_\mu=10$~GeV and (b)
$E_\mu=50$~GeV. The oscillation parameters used are given in
Eq.~(\ref{eq:params}).}
\end{figure}

Next we show in Fig.~\ref{fig:fixedE-vs-L}a the $L$ dependence of the
oscillations at a fixed muon energy of 10~GeV. Here we see
that $L\simeq2900$~km is the first minimum in the
survival probability and that $L\simeq5500$~km corresponds to a 
maximum in the survival probability;
matter effects are sizable for $L\sim5000-8000$~km.
In the appearance channels, matter effects 
turn on as $L$ increases, causing one of the appearance channels to
be highly suppressed.

Similar results for $E_\mu=50$~GeV are shown in
Fig.~\ref{fig:fixedE-vs-L}b. We see that the appearance event rates
scale roughly with $E_\mu$, as expected from the product of the $E_\mu^3$
behavior of the
unoscillated rates (Eq.~(\ref{eq:rate})) and the $E_\nu^{-2}$ dependence
of the oscillatory factor for small oscillation arguments
(Eqs.~\ref{eq:probs}--\ref{eq:arg0}). The $L$ dependence is relatively flat at
low $L$
for the channel that is not suppressed by matter, as expected from the product
of the
$1/L^2$ behavior of the flux and the $L^2$ dependence of the oscillatory
factor. The survival rates are much higher for $50$~GeV than for
$10$~GeV, but they do not reach a minimum for $L <
10000$~km.

\begin{figure}

\centering\leavevmode
\epsfxsize=6in\epsffile{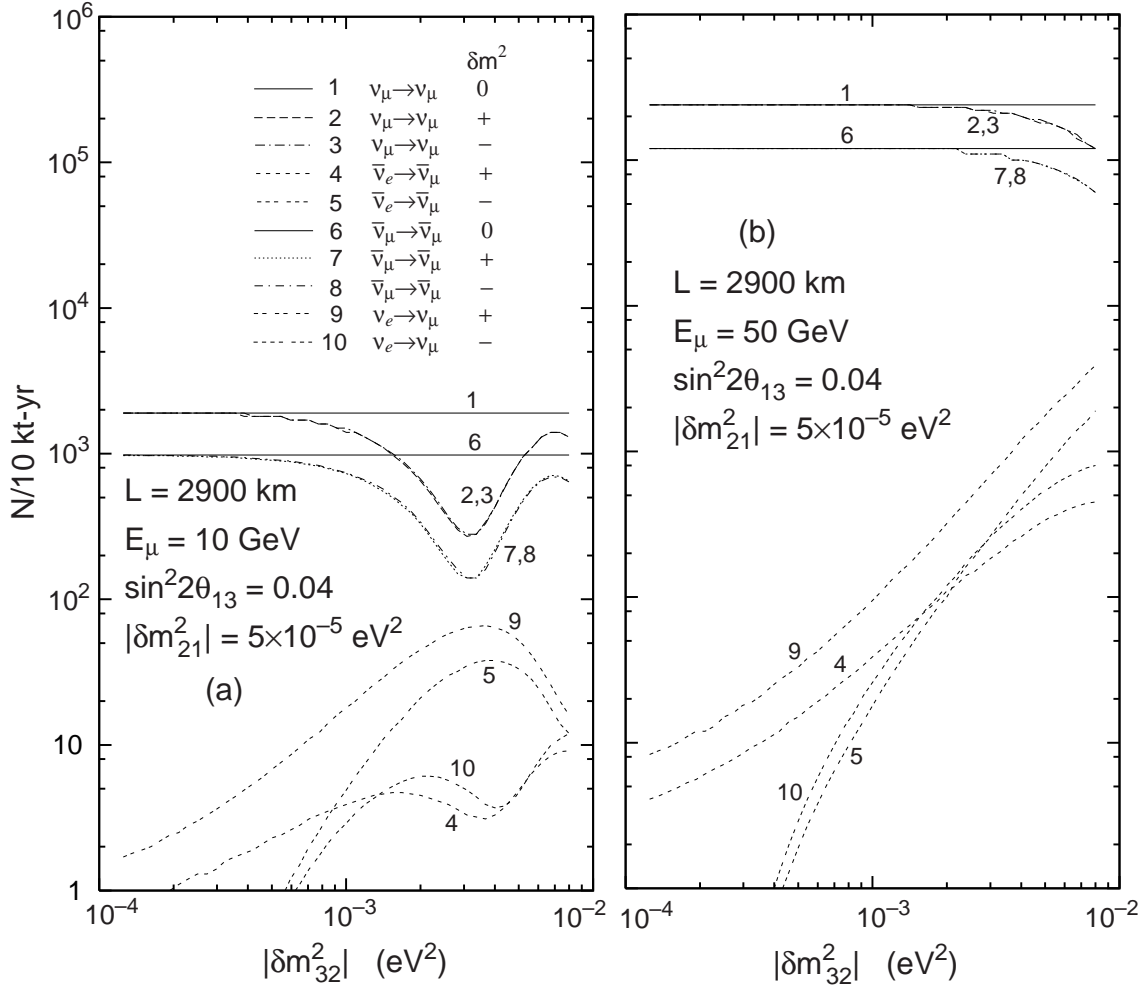}
\medskip
\caption[]{ \label{fig:2900-vs-dm2} Muon survival and appearance rates
per 10~kt-yr versus $|\delta m^2_{32}|$ at $L=2900$~km for (a) $E_\mu =
10$~GeV and (b) $E_\mu = 50$~GeV. The other oscillation parameters are
given in Eq.~(\ref{eq:params}).}
\end{figure}

Figures~\ref{fig:732-vs-E}--\ref{fig:fixedE-vs-L} assume $|\delta
m^2_{32}| = 3.5\times10^{-3}$~eV$^2$, the favored value from atmospheric
measurements from SuperK. Long-baseline experiments should be able to
determine $|\delta m_{32}^2|$ with high precision if the event rates associated
with oscillations are appreciable. Figure~\ref{fig:2900-vs-dm2} shows the
sensitivity of the event rates to changes in $\delta m_{32}^2$ for $L=2900$~km
and $E_\mu = 10$ or $50$~GeV. 
At $E_\mu = 10$~GeV the survival probability
has a minimum near $|\delta m^2_{32}| =
3.5\times10^{-3}$~eV$^2$, but the appearance rates are modest. 
At $E_\mu = 50$~GeV the appearance rates are much
higher, but the survival probability has a minimum 
only for higher $|\delta m^2_{32}|$.  
Therefore we see that at $L=2900$~km, 
if $\delta m^2_{32}$ is close to the value favored by SuperK, 
there is a trade-off between choosing a relatively low $E_\mu$ to 
optimize the survival probability measurements, or choosing a higher energy 
to optimize the appearance signal. 

\begin{figure}

\centering\leavevmode
\epsfxsize=6in\epsffile{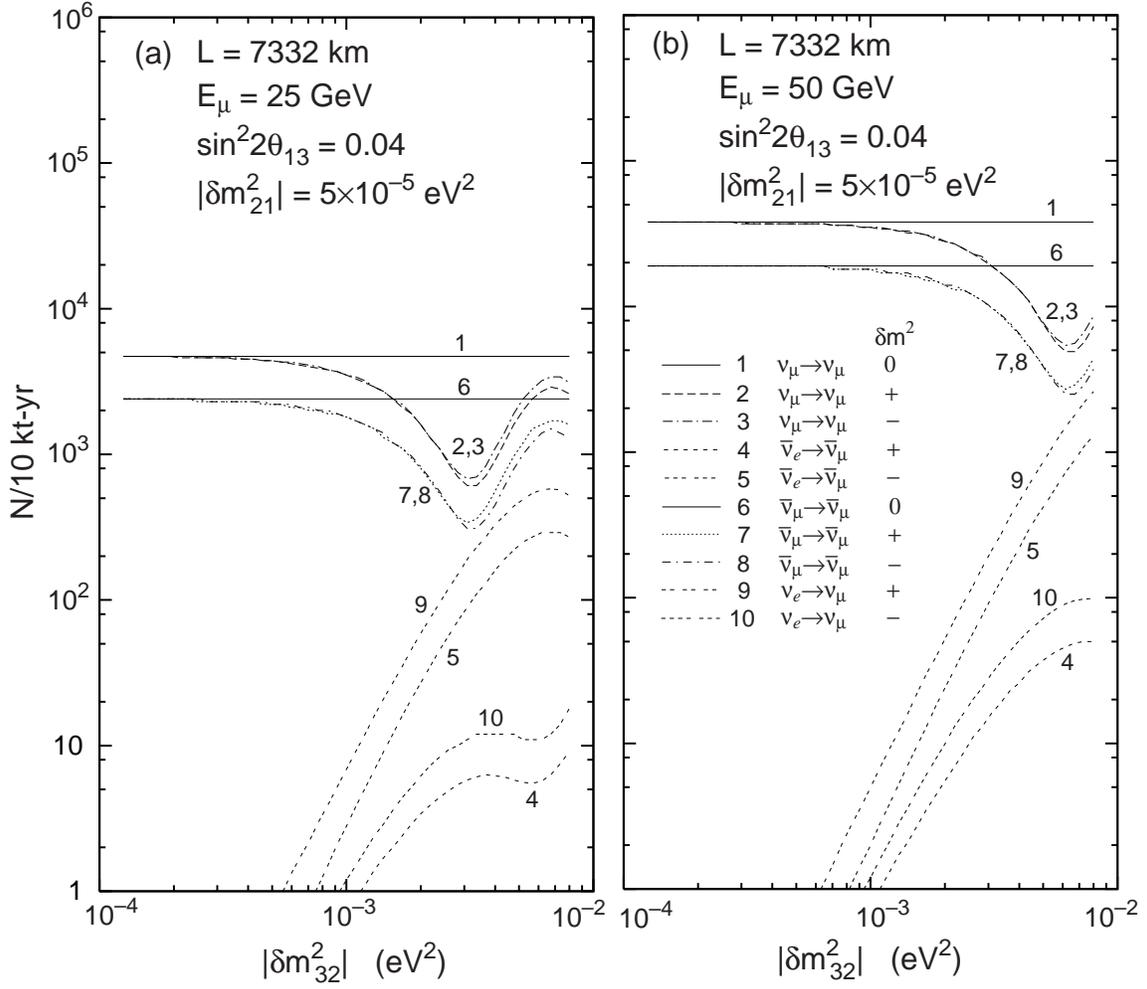}
\caption[]{ \label{fig:7332-vs-dm2} Same as Fig.~\ref{fig:2900-vs-dm2}
except $L=7332$~km.}
\end{figure}

Figure~\ref{fig:7332-vs-dm2} shows similar curves for $L=7332$~km with
$E_\mu = 25$ and 50~GeV. At $E_\mu = 25$~GeV the survival probability
has a minimum near $|\delta m^2_{32}| =
3.5\times10^{-3}$~eV$^2$,
with appreciable appearance rates for nearby values of $\delta m^2_{32}$,
{\it and} matter effects are evident for $|\delta
m^2_{32}| \ge 3\times10^{-3}$~eV$^2$. At $E_\mu = 50$~GeV, the
appearance rates are higher than for 25~GeV (especially in the
matter-suppressed channel) and the disappearance probability is
appreciable over a range of $|\delta m^2_{32}|$. However, matter effects
in the survival probability are noticeable only for $|\delta
m^2_{32}| > 7\times10^{-3}$~eV$^2$.

\begin{figure}

\centering\leavevmode
\epsfxsize=3.5in\epsffile{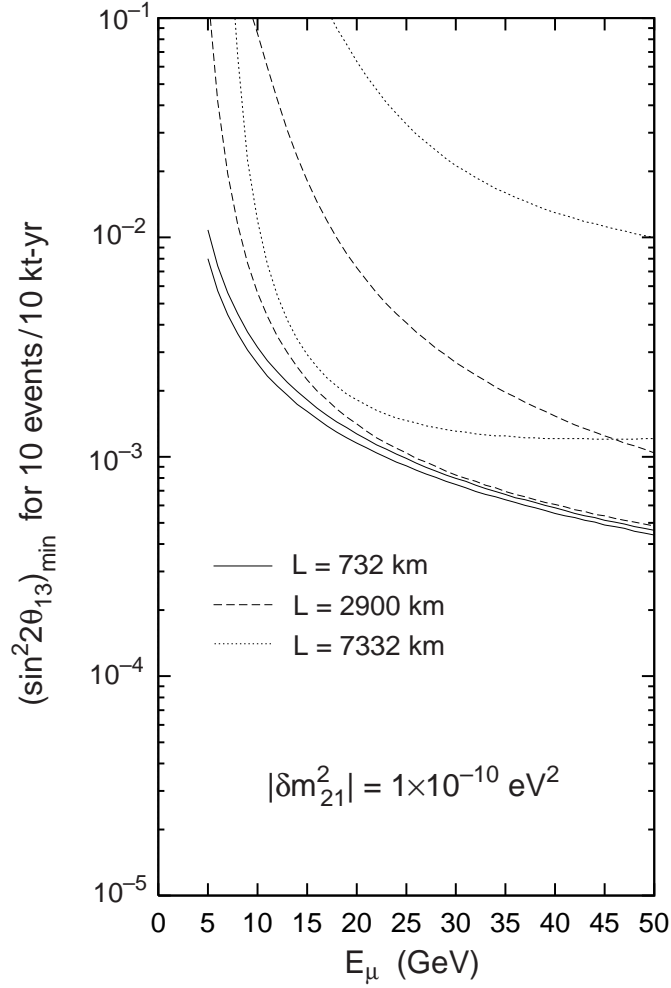}
\medskip
\caption[]{ \label{fig:minamp-vs-E} Minimum value of
$\sin^22\theta_{13}$ that gives 10 $\nu_e\to\nu_\mu$ appearance events
in a 10~kt detector with a source of $2\times10^{20}$ muon decays/yr, 
for $L =
732$~km (solid curves), $2900$~km (dashed curves), and $7332$~km (dotted
curves). The oscillation parameters are given in
Eq.~(\ref{eq:params}) and $\delta m^2_{21} \sim 1\times 10^{-10}$~eV$^2$.
The lower (upper) curve in each case is the result for $\delta
m^2_{32} > 0$ ($\delta m^2_{32} <0$). The corresponding curves for
$\bar\nu_e\to\bar\nu_\mu$ appearance are given approximately by
interchanging the upper and lower curves and dividing by a factor of 2.}
\end{figure}

Figure~\ref{fig:minamp-vs-E} shows the minimum value of
$\sin^2 2\theta_{13}$ that gives 10 $\nu_e\to\nu_\mu$ appearance events
versus $E_\mu$ for $L=732$, 2900, and 7332~km, when $\delta
m^2_{21} \sim 10^{-10}$~eV$^2$ (the solar vacuum oscillation value). The
lower (upper) curves are for matter-enhanced (matter-suppressed)
oscillations with $\delta m^2_{32} > 0$ ($\delta m^2_{32} < 0$); the
corresponding curves for $\bar\nu_e\to\bar\nu_\mu$ appearance can be
empirically obtained by changing the sign of $\delta
m^2_{32}$ and multiplying by 2. We see that increased $E_\mu$ improves the
ability to
discover appearance channels, although there are limiting returns for
going to very high $E_\mu$. The sensitivity to the matter-suppressed
channel is especially weak at longer distances (e.g. $L=7332$~km in
this case). The corresponding curves for the large-angle MSW solar
solution (e.g. $\delta m^2_{21} = 5\times10^{-5}$~eV$^2$) are much
different for $L=732$~km, and somewhat different at $L=2900$~km, due
to contamination by the subleading oscillation, as expected from the
discussion of Figs.~\ref{fig:732-vs-E} and ~\ref{fig:2900-vs-E}. If
large-angle solar oscillations are the correct solution to the solar anomaly,
then $L=7332$~km has the merit that the subleading oscillation does not
affect the interpretation of $\nu_e\to\nu_\mu$ appearance, as it does
at $L=732$ and 2900~km.

As discussed in Sec.~\ref{sec:leading-osc}, the leading oscillation
approximation leads to a simple expressions for the oscillation argument of the
$\nu_e\to\nu_\mu$ probability (see Eqs.~(\ref{eq:arg}) and
(\ref{eq:approxS})). Then the $\nu_e\to\nu_\mu$ appearance event rate
can be approximated by
\begin{eqnarray}
N &\simeq& \left< \Phi  P(\nu_e\to\nu_\mu) \sigma(\nu_\mu\to\mu^-)\right>
\label{eq:approxrate}\\
&\simeq& 300 \left( {n_0\over 2\times10^{20}\mu/{\rm yr}} \right)
\left({size\over 10 {\rm~kt}} \right)
\left({E_\mu\over10{\rm~GeV}} \right)^3
\left({7332{\rm~km}\over L} \right)^2
{\left< E_\nu \right>\over E_\mu}
\left< P(\nu_e\to\nu_\mu) \right> \,,
\nonumber
\end{eqnarray}
where the angle brackets denote averages over neutrino energy, and
$\left< E_\nu \right> = 0.6 E_\mu$. The average probability can be
estimated by evaluating the probability at $\left< E_\nu \right>$ in the
small $\sin^2 2\theta_{13}$ limit,
\begin{equation}
\left< P(\nu_e\to\nu_\mu) \right> \simeq {\sin^2 2\theta_{13}\over
\left|1-{\left< A \right>\over\delta m^2_{32}} \right|^2}
\sin^2 \left[ 1.27{\delta m^2_{32} L\over\left< E_\nu\right>}
\left| 1-{\left< A \right>\over\delta m^2_{32}} \right| \right]
\label{eq:approxP} \,,
\end{equation}
where $\left< A \right>$ is given by Eq.~(\ref{eq:A}) evaluated at
$\left< E_\nu \right>$. Equation~(\ref{eq:approxrate}) gives reasonably
accurate results whenever $\sin^2 2\theta_{13}$ is below 0.01 provided that
contributions to the subleading oscillation are not important. Approximate
antineutrino
rates can be obtained  by the substitution
$A \to -A$ in Eq.~(\ref{eq:approxP}) and by dividing the results of
Eq.~(\ref{eq:approxrate}) by~2.


\section{More Detailed Simulations}
\label{sec:detail}

To obtain a more realistic calculation of the event rates at a neutrino factory
we have simulated a neutrino beam formed by muons decaying along a 1~km long
straight-section. In our simulations the muon beam has a momentum spread given
by $\sigma_p/p = 0.02$, and horizontal and vertical beam divergences given by
$\sigma_\theta = 0.1 / \gamma$ where
$\gamma = E_\mu / m_\mu$. The finite beam momentum spread and angular
divergence modify the event rates at a distant site by a few percent. We have
also included in our simulations a parametrization of the detector
resolution functions for an iron-scintillator neutrino
detector~\cite{minos,thesis}. We assume that muon energies can be measured by
range with a precision given by $\sigma_E / E = 0.05$. Shower energies are
assumed to be measured with precisions given by $\sigma_E / E = 0.53 /
\sqrt{E}$ if $E > 3$~GeV and $\sigma_E / E = 0.80 / \sqrt{E}$ if $E < 3$~GeV
with, in both cases, a constant term of 0.07 added in quadrature.

\begin{figure}
\epsfxsize3.5in
\centerline{\epsffile{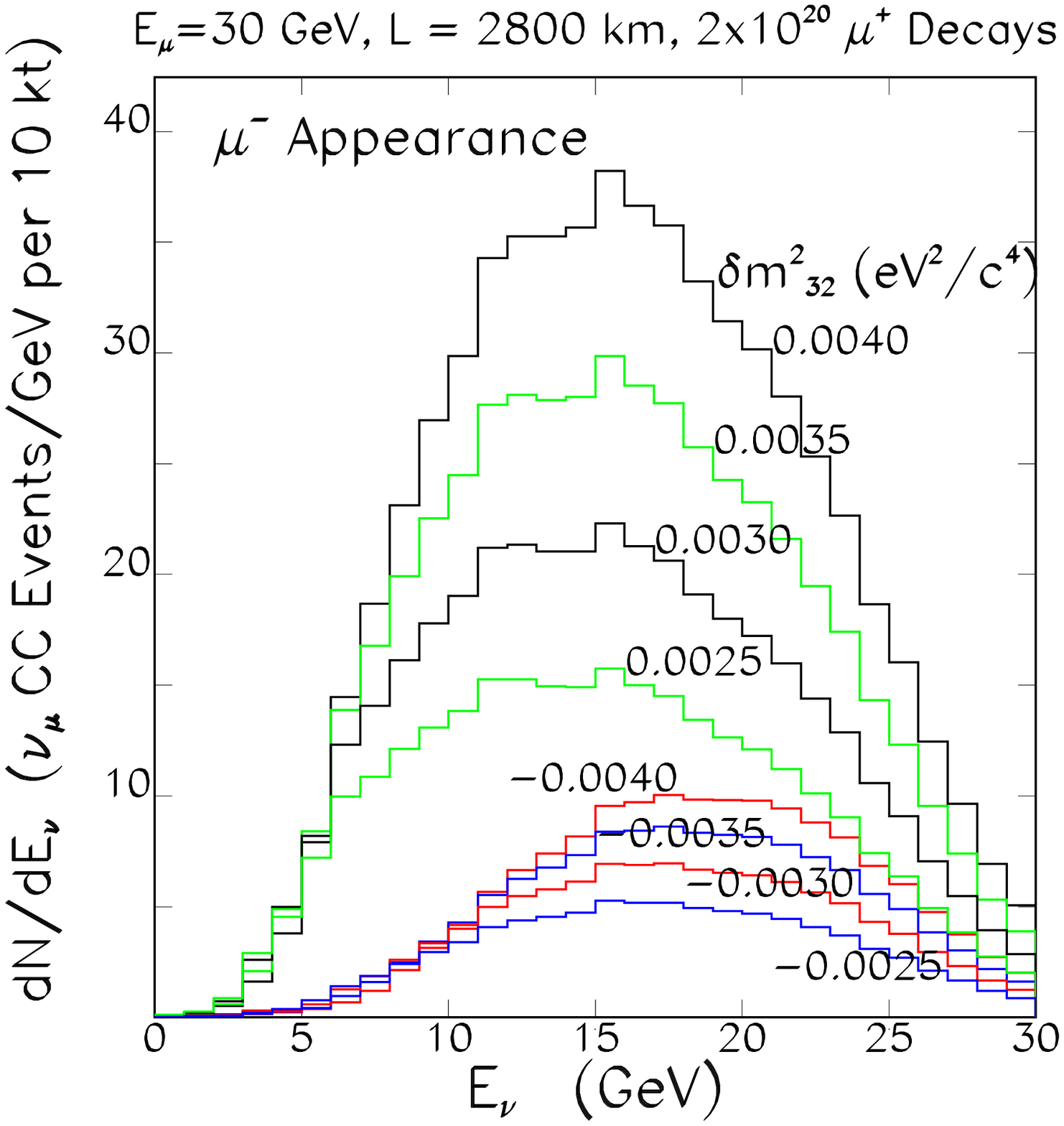}}
\bigskip
\caption[]{Predicted measured energy distributions for CC events tagged by a
wrong-sign (negative) muon from $\nu_e \rightarrow\nu_\mu$ oscillations, shown
for various $\delta m^2_{32}$, as labelled. The predictions correspond to $2
\times 10^{20}$ decays, $E_\mu = 30$~GeV, $L = 2800$~km, with the values for
$\delta m^2_{12}, s_{13}, s_{23}, s_{12}$, and $\delta$ given in
Eq.~(\ref{eq:params}).}\label{fig:ws_muon+}
\end{figure}

\begin{figure}
\epsfxsize3.5in
\centerline{\epsffile{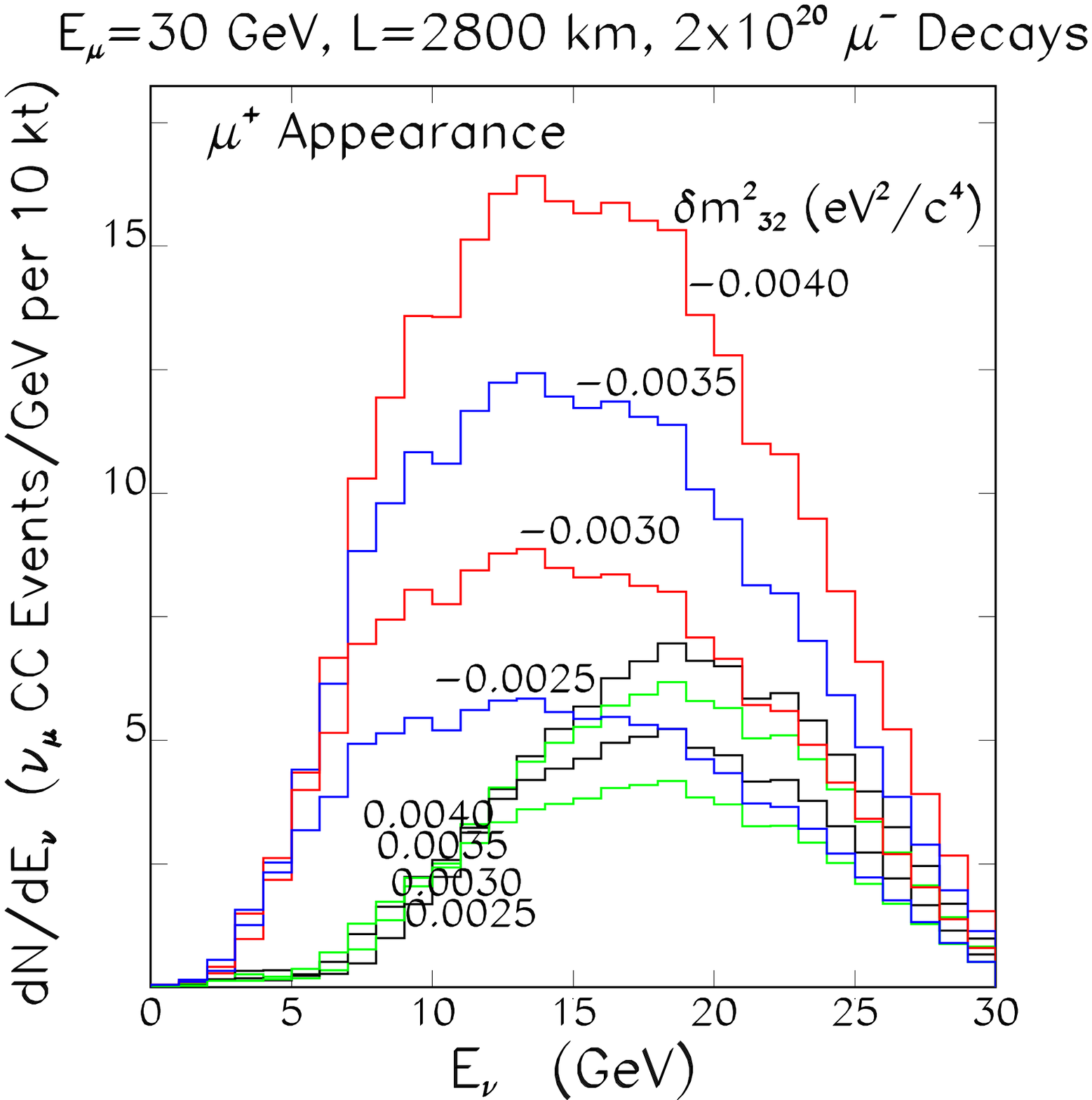}}
\bigskip
\caption[]{Predicted measured energy distributions for CC events
tagged by a wrong-sign (positive) muon from $\bar{\nu}_e \rightarrow
\bar{\nu}_\mu$ oscillations, shown for various $\delta m^2_{32}$, as
labelled. The predictions correspond to $2 \times 10^{20}$ decays,
$E_\mu = 30$~GeV, $L = 2800$~km, with the values for
$\delta m^2_{12}, s_{13}, s_{23}, s_{12}$, and $\delta$ given
in Eq.~(\ref{eq:params}).}\label{fig:ws_muon-}
\end{figure}

For each of the relevant oscillation channels, the predicted annual CC event
samples (from $2 \times 10^{20}$ muon decays) corresponding to the oscillation
parameters of Eq.~(\ref{eq:params})
are shown in Table~\ref{table:rates} for both signs of
$\delta m_{32}^2$. The tabulated rates are for a 10~kt detector at baselines of
$L=732$~km, 2800~km, and 7332~km, with $E_\mu =$ 10, 30 and 50~GeV. The
statistical uncertainties on the calculated event rates from the Monte Carlo
simulations are about 2.5\%.
Consider first the sensitivity of the oscillation signals to matter effects,
and hence to the sign of $\delta m_{32}^2$. Note that most of the event rates
are insensitive to matter effects. However, the $\nu_e \rightarrow \nu_\mu$ CC
rates are significantly modified by matter, and hence a measurement of
events with wrong-sign muons can in principle
determine the sign of $\delta m_{32}^2$. To illustrate this, consider the
$\nu_e \rightarrow \nu_\mu$ CC rates when L = 2800~km and $E_\mu = 50$~GeV
(30~GeV). If $\delta m_{32}^2$ is positive, 894 (486) wrong-sign muon events
are expected in a 10~kt-yr data sample. However, if $\delta m_{32}^2$ is
negative, the corresponding rates are 338 (130) events. Hence, if all the
oscillation parameters are known except the sign of $\delta m_{32}^2$, the
event rates can be used to determine the sign. The statistical significance of
this sign determination improves slowly with increasing $E_\mu$. In practice,
the oscillation parameters will not be precisely known. However, the sign of
$\delta m_{32}^2$ can still be uniquely determined from the wrong-sign muon
measurements if both $\nu_e \rightarrow \nu_\mu$ rates (positive stored muons)
and $\bar{\nu}_e \rightarrow \bar{\nu}_\mu$ rates (negative stored muons) are
determined. For the examples we are considering, if $\delta
m_{32}^2$ is positive the $\nu_e \rightarrow \nu_\mu$ rates will be
larger than the $\bar{\nu}_e \rightarrow \bar{\nu}_\mu$ rates by a factor of a
few. In contrast, if $\delta m_{32}^2$ is negative the $\nu_e \rightarrow
\nu_\mu$ rates will be comparable to, or smaller than, the $\bar{\nu}_e
\rightarrow \bar{\nu}_\mu$ rates.
There is additional information in the spectrum of CC events producing
wrong-sign muons. The predicted measured energy distributions (including
detector resolutions), for CC events containing wrong-sign muons are shown as a
function of both the magnitude and sign of $\delta m_{32}^2$ in
Figs.~\ref{fig:ws_muon+} and \ref{fig:ws_muon-} when, respectively, 30~GeV
positive and negative muons are stored in the ring, and $L = 2800$~km. Both the
shapes and normalizations of the distributions are sensitive to the sign and
magnitude of $\delta m_{32}^2$. In particular if $\delta m_{32}^2$ is negative
(positive) the $\mu^-$ appearance events from $\nu_e \rightarrow \nu_\mu$
oscillations will peak at higher (lower) energies than the $\mu^+$ appearance
events from $\bar{\nu}_e \rightarrow \bar{\nu}_\mu$ oscillations. We conclude
that if both positive and negative muons can be stored at different times in
the neutrino factory, then wrong-sign muon appearance measurements can
distinguish the sign of $\delta m_{32}^2$ and determine its magnitude.

\begin{figure}
\epsfxsize3.5in
\centerline{\epsffile{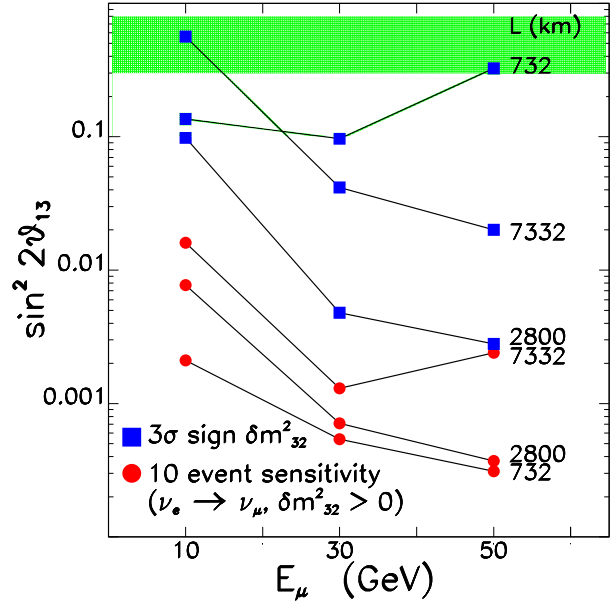}}
\bigskip
\caption[]{The value of $\sin^2 2\theta_{13}$ that yields, in a 10~kt detector
at $L = 2800$~km, (a) 10 events per $2 \times 10^{20} \mu^+$ decays (boxes), and
(b) a three standard deviation determination of the 
 sign of $\delta m^2_{32}$ (circles) based on the ratio of wrong-sign
 muon rates when alternately $2 \times 10^{20}$ $\mu^+$ and 
 $2 \times 10^{20}$ $\mu^-$ decay in the neutrino beam-forming 
 straight section. The $\sin^2 2\theta_{13}$ sensitivity is shown versus $E_\mu$ and
$L$ (as labelled). The calculations assume the values for $|\delta m^2_{32}|$,
$\delta m^2_{12}, s_{23}, s_{12}$, $\delta$ given
in Eq.~(\ref{eq:params}). The shaded region is excluded by the existing data.}
\label{fig:sensitivity}
\end{figure}

We next consider a less favorable region of parameter space in which the $\nu_e
\rightarrow \nu_\mu$ oscillation amplitude is reduced by a large factor.
Table~\ref{table:s_1} presents the event rates when $\sin^22\theta_{13}$ 
is reduced by an
order of magnitude ($s_{13} = 0.032$ or $\sin^22\theta_{13} = 0.004$)
with the other oscillation parameters as in Eq.~(\ref{eq:params}).
Returning to our example ($L = 2800$~km, $E_\mu = 50$~GeV) we note that if
$\delta m_{32}^2$ is positive (negative) we expect 108 (25) wrong-sign CC
events per 10~kt-yr when positive muons are stored, and 29 (30) events when
negative muons are stored.
Hence, in this case we are still above, but close to, the threshold below which
a statistically significant determination of the sign of $\delta m_{32}^2$ will
only be possible with larger data samples.
Based on Tables~\ref{table:rates} and \ref{table:s_1} we can estimate the
minimum values of $\sin^22\theta_{13}$ 
(when all of the other parameters are as specified in Eq.~(\ref{eq:params}))
for which (i) we expect a measurable wrong-sign muon signal (10~event
sensitivity), and (ii) from the measured ratio of wrong-sign 
 muon rates obtained when positive and negative muons 
 are alternately stored we can determine the sign of $\delta m_{32}^2$ (at 3
standard deviations). Figure~\ref{fig:sensitivity} summarizes our estimates for
these minimum $\sin^22\theta_{13}$ 
versus $L$ and $E_\mu$. Note that to measure a
wrong-sign muon signal over the largest $\sin^22\theta_{13}$ range, high $E_\mu$
and ``short" $L$ are preferred. However, if $L$ is too short the unoscillated
CC event rates get very large, and hence backgrounds become significant. If the
background rates are at the level of $10^{-5}$--$10^{-6}$ of the total CC
rates~\cite{cadenas}, then $L = 2800$~km is probably preferred over 732~km. To
be able to determine the sign of $\delta m_{32}^2$ over the largest 
$\sin^22\theta_{13}$ range,
high $E_\mu$ is once again preferred. Since matter effects become small for
short $L$ and event statistics become small for very large $L$, there
appears to be an optimal $L$ for determining the sign of $\delta m_{32}^2$. 
Of the 3 baselines we have considered, $L =
2800$~km is preferred. From these considerations we would conclude that for
wrong-sign muon measurements $L = 2800$~km is a good choice for the baseline,
and high energy stored muons ($E_\mu = 50$~GeV) are preferred, although we note
that decreasing $E_\mu$ from 50~GeV to 30~GeV decreases the sensitive 
$\sin^22\theta_{13}$ range by less than a factor of two.

\begin{figure}
\epsfxsize3.15in
\centerline{\epsffile{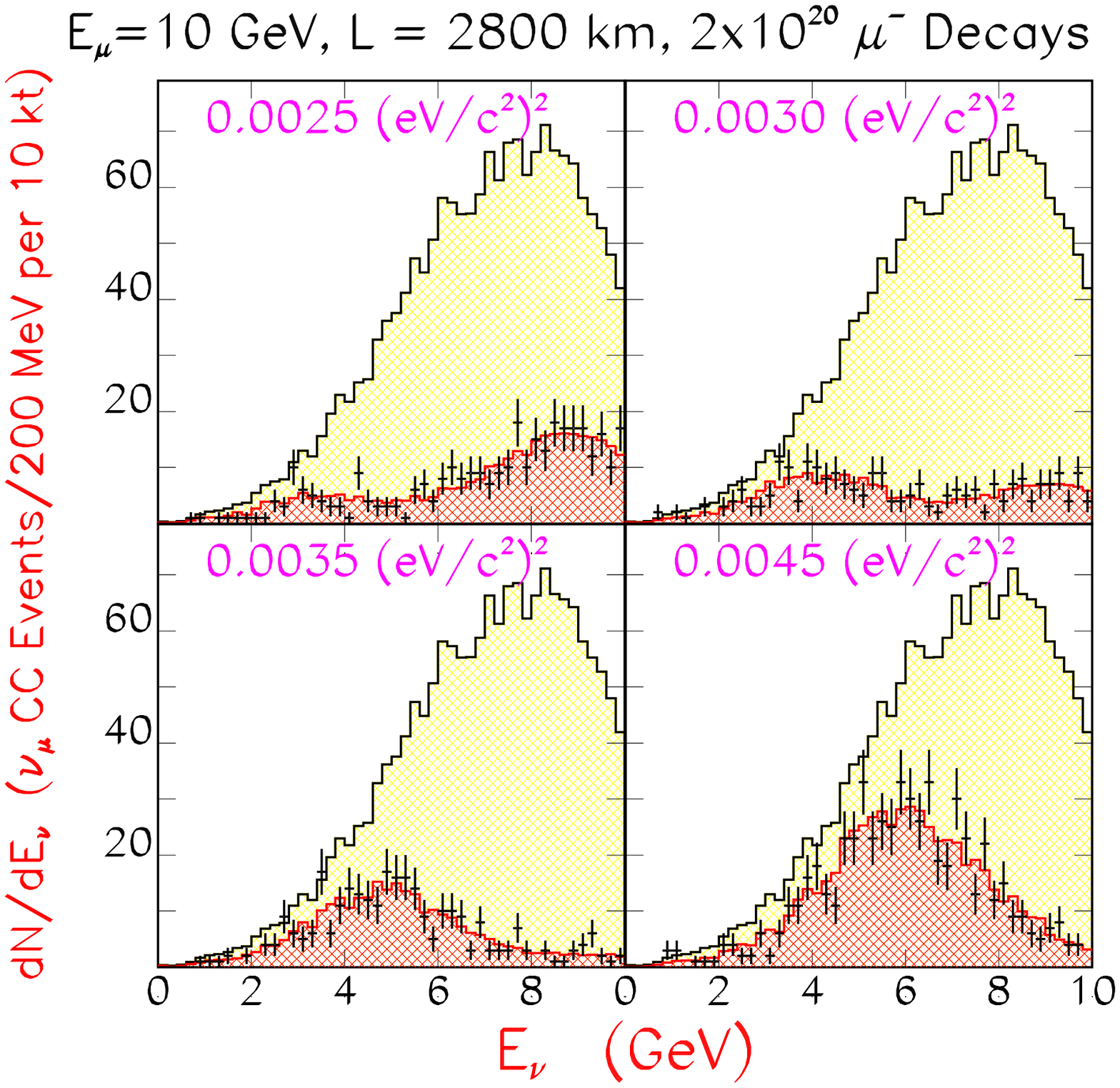}}
\medskip
\caption[]{Predicted measured energy distributions for CC $\nu_\mu \rightarrow
\nu_\mu$ events shown for four different $\delta m^2_{32}$ (darkly shaded
distributions) as labelled. The predictions correspond to $2 \times 10^{20}$
decays, $E_\mu = 10$~GeV, $L = 2800$~km, with the values for
$\delta m^2_{12}, s_{13}, s_{23}, s_{12}$, and $\delta$ given
in Eq.~(\ref{eq:params}).
The predicted distribution has been used to generate a Monte Carlo dataset 
with the statistics corresponding to a 10~kt-yr dataset (points with error bars).
The lightly shaded histograms show the predicted distributions in the absence
of oscillations.}\label{fig:10gev_disap}
\end{figure}

\begin{figure}
\epsfxsize3.15in
\centerline{\epsffile{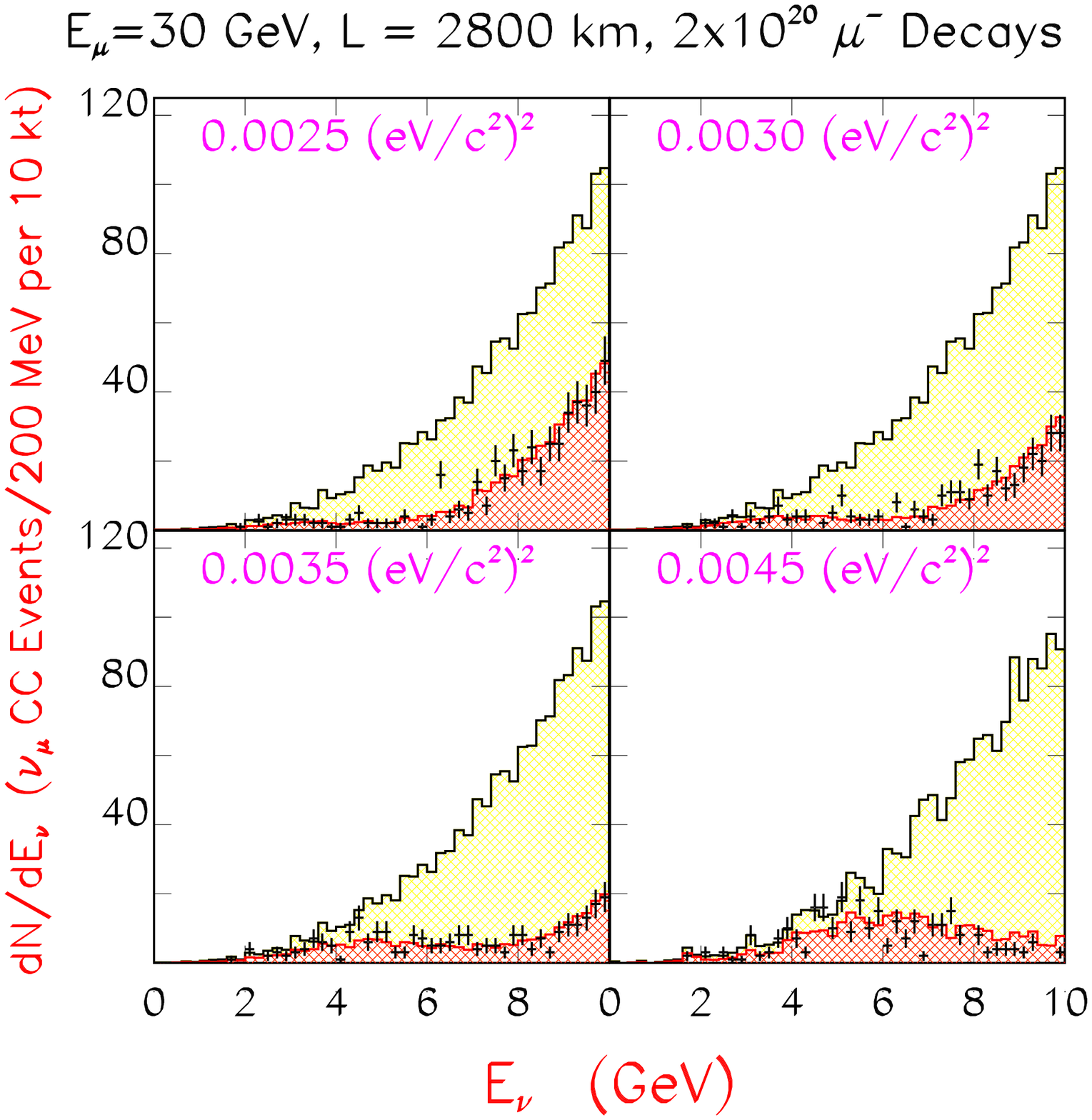}}
\bigskip
\caption[]{Predicted measured energy distributions for CC
$\nu_\mu \rightarrow \nu_\mu$ events shown for four different
$\delta m^2_{32}$ (darkly shaded distributions) as labelled.
The predictions correspond to $2 \times 10^{20}$ decays,
$E_\mu = 30$~GeV, $L = 2800$~km, with the values for
$\delta m^2_{12}, s_{13}, s_{23}, s_{12}$, and $\delta$ given
in Eq.~(\ref{eq:params}).
The predicted distribution has been used to generate a Monte Carlo dataset
with the statistics corresponding to a 10~kt-yr dataset (points with error bars).
The lightly shaded histograms show the predicted distributions in the absence
of oscillations.}\label{fig:30gev_disap}
\end{figure}

\begin{figure}
\epsfxsize5.5in
\centerline{\epsffile{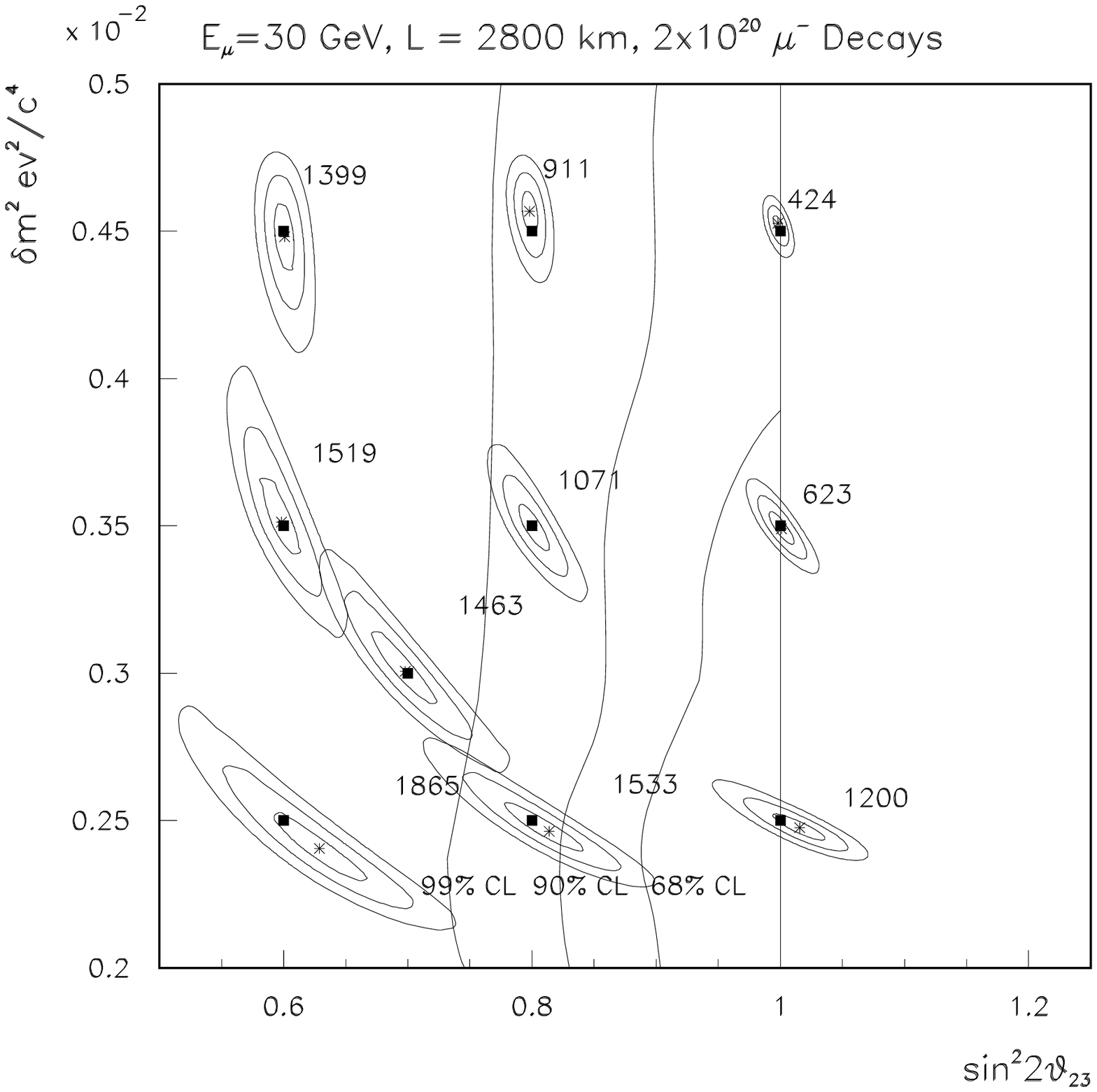}}
\bigskip
\caption[]{ \label{fig:30gev_disap_fit}
Fit to muon neutrino survival distribution for $E_\mu=30$ GeV and $L=2800$~km for 10
pairs of sin$^2 2\theta$, $\delta m^2$ values. For each fit, the
1$\sigma$,\ 2$\sigma$
and 3$\sigma$ contours are shown. The generated points are indicated by the
dark
rectangles and the fitted values by stars. The SuperK 68\%, 90\%, and 95\% 
confidence
levels are superimposed. Each point is labelled by the predicted number of 
signal events for that point.}
\end{figure}

\begin{figure}
\epsfxsize5.5in
\centerline{\epsffile{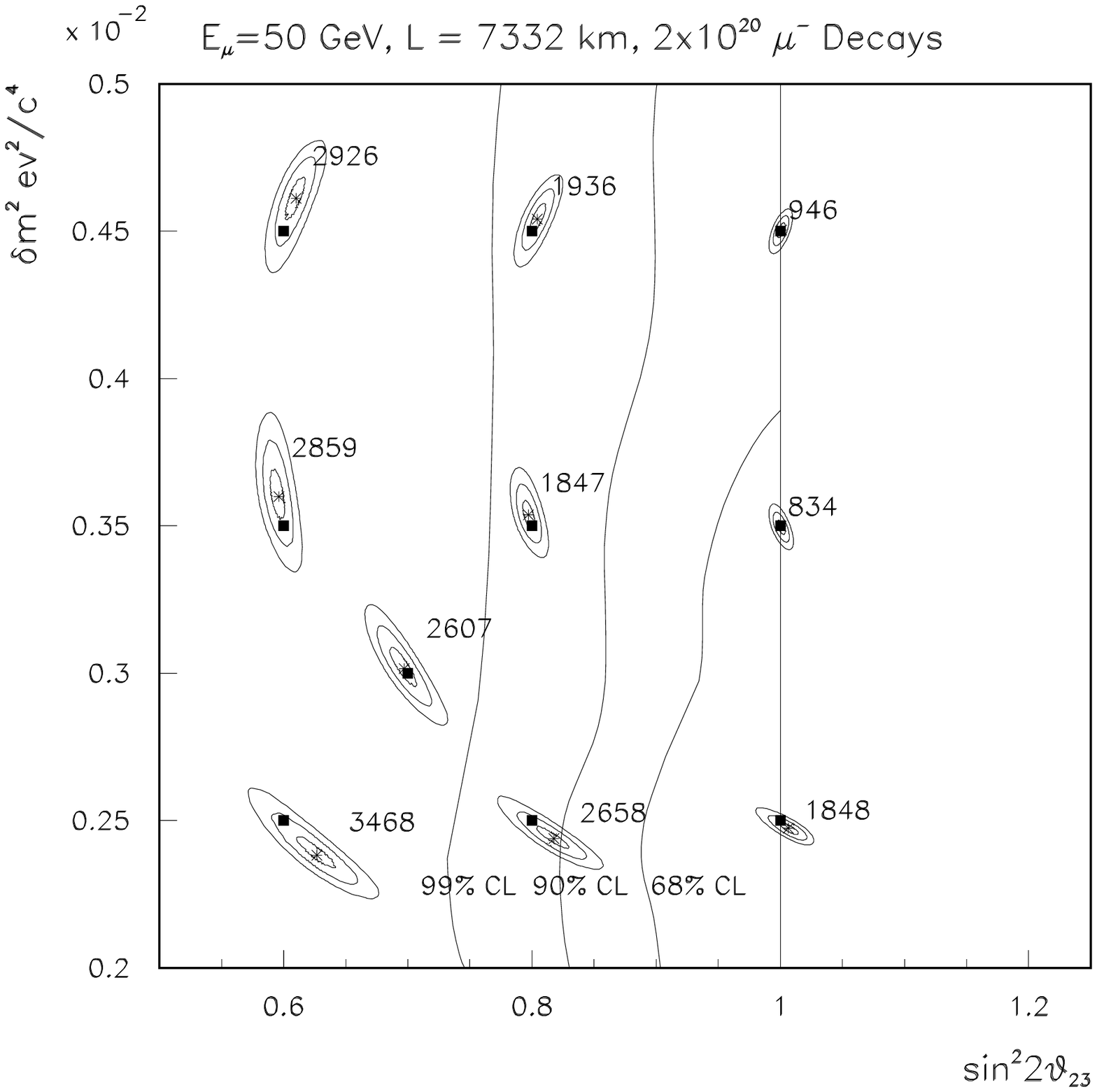}}
\bigskip
\caption[]{ \label{fig:50gev_disap_fit}
Fit to muon neutrino survival distribution for $E_\mu=50$ GeV and $L=7332$~km for 10
pairs of $\sin^2 2\theta$, $\delta m^2$ values. For each fit, the
1$\sigma$,\ 2$\sigma$
 and 3$\sigma$ contours are shown. The generated points are indicated by the
dark
rectangles and the fitted values by stars. The SuperK 68\%, 90\%, and 95\% 
confidence
levels are superimposed. Each point is labelled by the predicted number of
signal events for that point.}
\end{figure}

We next turn our attention to muon survival measurements.
The $\nu_\mu \rightarrow \nu_\mu$ rates shown in Table~\ref{table:rates} are
significantly less than the corresponding rates in the absence of oscillations.
As an example, consider $L = 2800$~km with $E_\mu = 10$~GeV and 30~GeV. The
predicted measured energy distributions (including detector resolutions) for
interacting muon neutrinos are shown for the two energies in
Figs.~\ref{fig:10gev_disap} and \ref{fig:30gev_disap} respectively versus
$\delta m_{32}^2$ with the other oscillation parameters given by
Eq.~(\ref{eq:params}). The shapes and normalizations for the predicted
distributions are very different from the expected distribution that would be
observed in the absence of oscillations. The shapes of the predicted $\nu_\mu$
CC interaction distributions are sensitive to  the magnitude of $\delta
m_{32}^2$. The dip in the predicted rate that corresponds to the first
oscillation maximum (when $1.267\;\delta m_{32}^2 L/E_\nu = \pi$) can be
clearly seen. At the higher $E_\mu$ the contribution from poorly measured
higher energy $\nu_\mu$ CC interactions reduces the significance of the dip. To
understand the statistical precision with which $\delta m_{32}^2$ and 
$\sin^22\theta_{13}$
can be extracted from a fit to the measured $\nu_\mu$ CC interaction
distribution, we have generated Monte Carlo data sets with the appropriate
statistics for various values of the oscillation parameters, and fit the
resulting simulated distributions.
As an example, Figs.~\ref{fig:30gev_disap_fit} and \ref{fig:50gev_disap_fit}
show fit results for $E_\mu = 30$~GeV with $L = 2800$~km
and $E_\mu=50$~GeV with $L = 7332$~km.
The precision of the fits for various $L$ and $E_\mu$
are summarized in Table~\ref{table:limits} for  $\sin^22\theta_{23}=1.0$
 and $\delta m^2_{32}=0.35\times10^2{\rm\,eV}^2$.
In order to extract the fitted information,
we have had to limit the fits to energy ranges where the information is
maximally
available. It can be seen from  Table~\ref{table:limits} that the parameters
are ill
constrained for the shorter baseline of 732 km. For the 10~GeV ring,
$L=2800$~km and for the
 30 or 50~GeV ring $L=7332$~km seem to be the optimal baseline lengths for
maximum
 precision in sin$^22\theta_{23}$ and  $\delta m^2_{32}$.

\section{Conclusions}

Within the framework of three-flavor oscillations, we have explored the
sensitivity of muon appearance and muon disappearance measurements at a
neutrino factory in which there are $2 \times 10^{20}$ muon decays per
year in a beam-forming straight section that points at a 10~kt detector.
Three stored muon energies
(10, 30, and 50~GeV) and three baselines (732, 2800, and 7332~km) have
been explicitly considered. Our results are summarized in Table~VI.

If data can be taken alternately with
positive and negative muons stored in the ring, a measurement of the
wrong-sign muon appearance rates and spectra can uniquely determine
the sign of $\delta m^2_{32}$ provided the oscillation amplitude is
sufficiently large (sin$^22\theta_{13} > 0.005$ for $L = 2800$~km
and $E_\mu = 30$~GeV).
To demonstrate this, we have considered determining 
   the sign of $\delta m^2_{32}$ by measuring the ratio of wrong-sign 
   muon rates when alternately positive and negative muons are 
   stored.
Of the three baselines we have considered, $L = 2800$~km is preferred
for this measurement since
$L = 732$~km is too short
(matter effects too small) and $L = 7332$~km is too long (statistics
too limited) to obtain good sensitivity to the sign of $\delta m^2_{32}$.
We note that a global fit to both wrong-sign muon event rates 
   and energy distributions with $\delta m^2_{32}$ (including its sign) 
   left as a free parameter would improve the sensitivity to the sign 
   of $\delta m^2_{32}$, and may result in a different preferred baseline 
   choice~\cite{BGRW3}.
The sensitivity to wrong-sign muon appearance, characterized
by the minimum sin$^22\theta_{13}$ for which a signal can be detected
at the 10~events per year level, improves linearly with $E_\mu$. For
$L = 2800$~km and $E_\mu = 30$~GeV the minimum sin$^22\theta_{13} = 0.0007$.
This can be improved to sin$^22\theta_{13} = 0.0004$
by either increasing $E_\mu$ to 50~GeV, or by increasing the detector mass
by a factor of 1.8.

The sensitivity for the survival measurements can be optimized
by choosing $L$ and $E_\mu$ so that $L/E_\nu$ is close to the first
minimum in the survival probability. For $L = 2800$~km this corresponds
to $E_\mu \sim 10$~GeV. For $\delta m^2_{32} = 0.0035$~eV$^2/$c$^4$
and sin$^22\theta_{23} = 1$
the statistical precision of the resulting $\delta m^2_{32}$ and
sin$^22\theta_{23}$ measurements based on fitting the observed
$\nu_\mu \rightarrow \nu_\mu$ spectra are respectively 2.4\% and 1.1\%.
However, optimization for the appearance channels
suggests choosing a higher $E_\mu$. For example, choosing $E_\mu = 30$~GeV
the precisions for the $\delta m^2_{32}$ and
sin$^22\theta_{23}$ measurements become 3.2\% and 2.0\%. With these levels
of statistical precision, systematic effects
(for example, the uncertainty on the neutrino
flux) may dominate.
For the region of three-flavor-mixing parameter space that we have explored,
we conclude that $L = 2800$~km with $E_\mu = 30$~GeV would enable the
very precise determination of $|\delta m^2_{32}|$ and
$\sin^2 2\theta_{23}$ from muon survival, the determination of
$\sin^2 2\theta_{13}$
from muon appearance, and the sign of $\delta m^2_{32}$ from matter
effects (e.g., by comparing $\nu_e \to \nu_\mu$ with $\bar\nu_e \to
\bar\nu_\mu$).

The above analysis assumes that $CP$ is conserved, i.e.,
$\delta = 0$. $CP$ violation may be important whenever the
effects of the subleading mass scale $\delta m^2_{21}$ are
appreciable, e.g., at short and intermediate
distances~\cite{CP}. A detailed analysis of the $CP$-violating
case will be given elsewhere~\cite{BGRW3}.

Finally, we note that if no appearance signal were observed, the implied very
low value of sin$^22\theta_{13}$ might provide a window of opportunity for
detecting oscillations driven by the smaller (solar neutrino deficit) scale
$\delta m^2_{21}$, should it be the large-angle MSW solution. This requires
further study.


\acknowledgments
We thank S.~Parke for discussions.
This research was supported in part by the U.S. Department of Energy under
Grants No.~DE-FG02-94ER40817 and No.~DE-FG02-95ER40896  and in part by the
University of Wisconsin Research Committee with funds granted by the Wisconsin
 Alumni Research Foundation. Part of this work was performed at the Fermi
National Accelerator Laboratory, which is operated by the Universities
 Research Association, under contract No.~DE-AC02-76CH03000 with the
 U.S.~Department of Energy.


\begin{table}
\squeezetable
\caption[]{Some candidate
neutrino factory accelerator sites and long-baseline detector sites.}
\label{table:baselines}
\begin{tabular}{lcccccccccccc}
&\multicolumn{12}{c}{Sources}\\
\cline{2-13}
&\multicolumn{3}{c}{FNAL}& \multicolumn{3}{c}{BNL}&
\multicolumn{3}{c}{CERN}& \multicolumn{3}{c}{KEK}\\
Targets & $L$\,(km)& $\left< Y_e\rho \right>$&
$\vcenter{\hbox{\ dip}\hbox{angle}}\, ({}^\circ$)&
$L$\,(km)& $\left< Y_e\rho \right>$&
$\vcenter{\hbox{\ dip}\hbox{angle}}\, ({}^\circ$)&
$L$\,(km)& $\left< Y_e\rho \right>$&
$\vcenter{\hbox{\ dip}\hbox{angle}}\, ({}^\circ$)&
$L$\,(km)& $\left< Y_e\rho \right>$&
$\vcenter{\hbox{\ dip}\hbox{angle}}\, ({}^\circ$)\\
\hline
MINOS& 732& 1.67& 3.3& 1715& 1.67& 7.7& 6596& 2.01& 31& 8559& 2.20& 42\\
BNL& 1289& 1.67& 5.8& &&& 5921& 1.91& 28& 9607& 2.31& 49\\
SLAC& 2913& 1.67& 13& 4513& 1.69& 19& 8590& 2.20& 42& 7720& 2.12& 37\\
Gran Sasso& 7332& 2.09& 35& 6565& 2.00& 31& 743& 1.67& 3.3& 8844& 2.23& 44\\
Kamioka& 9165& 2.26& 46& 9667& 2.31& 49& 8774&  2.22& 44& 252& 1.67& 1.1
\end{tabular}
\end{table}

\bigskip

\begin{table}
\caption[]{Signatures for various oscillation channels when negative
muons are stored.}
\label{tab:osc-probs}
\begin{tabular}{lccc}
 & & produced& detected\\
signature& oscillation& lepton& muon\\
\hline
$\nu_\mu$-survival & $\nu_\mu\to\nu_\mu$& $\mu^-$& $\mu^-$\\
$\nu_\tau$-appearance & $\nu_\mu\to\nu_\tau$& $\tau^-$& $\mu^-$\\
$\bar\nu_\mu$-appearance& $\bar\nu_e\to\bar\nu_\mu$& $\mu^+$& $\mu^+$\\
$\bar\nu_\tau$-appearance& $\bar\nu_e\to\bar\nu_\tau$& $\tau^+$& $\mu^+$
\end{tabular}
\end{table}

\begin{table}
\caption[]{Predicted event rates in a 10~kt detector
per $2 \times 10^{20}$ muon decays, with the oscillation
parameters specified by Eq.~(\ref{eq:params}).}
\label{table:rates}
\smallskip
\begin{tabular}{cccccccc}
& sign of& \multicolumn{3}{c}{$E_{\mu^+}$ (GeV)}&
\multicolumn{3}{c}{$E_{\mu^-}$ (GeV)}\\
& $\delta m^2_{32}$& 10& 30& 50& 10& 30& 50\\
\hline
\multicolumn{8}{c}{$L=732$ km}\\
$\nu_\mu$ (no osc)& & 14,300& 382,000& 1,780,000& 29,000& 772,000& 3,560,000\\
$\nu_\mu\to\nu_\mu$& $+$& 11,200& 372,000& 1,760,000& 22,800& 750,000&
3,520,000\\
&  $-$& 11,200& 372,000& 1,760,000& 22,900& 751,000& 3,530,000\\
$\nu_e$ (no osc)& & 24,200& 656,000& 3,050,000& 12,400& 329,000& 1,520,000\\
$\nu_e\to\nu_\mu$& $+$& 158& 404& 1000& 69.4& 276& 486\\
& $-$& 106& 410& 718& 65.2& 223& 380\\
\hline
\multicolumn{8}{c}{$L=2800$ km}\\
$\nu_\mu$ (no osc)& & 1140& 27,500& 123,000& 2200& 52,500& 244,000\\
$\nu_\mu\to\nu_\mu$& $+$& 168& 18,800& 107,000& 274& 36,000& 212,000\\
& $-$& 164& 18,900& 107,000& 274& 36,200& 212,000\\
$\nu_e$ (no osc)& & 1400& 43,300& 211,000& 900& 22,600& 102,000\\
$\nu_e\to\nu_\mu$& $+$& 54& 486& 894& 4& 88& 224\\
& $-$& 3.4& 130& 338& 36& 214& 356\\
\hline
\multicolumn{8}{c}{$L=7332$ km}\\
$\nu_\mu$ (no osc)& & 152& 3880& 17,500& 294& 7660& 35,600\\
$\nu_\mu\to\nu_\mu$& $+$& 84& 592& 7220& 130& 1040 & 14,900\\
& $-$& 72& 558& 7340& 160& 1130& 15,000\\
$\nu_e$ (no osc)& & 21.8& 6680 & 30,500& 120& 3300& 15,200\\
$\nu_e\to\nu_\mu$& $+$& 21.6& 274& 300& 0.66& 9& 19.8\\
& $-$& 1.14& 18& 38& 12.8& 132& 145
\end{tabular}
\end{table}

\begin{table}
\caption[]{Predicted event rates in a 10~kt detector
per $2 \times 10^{20}$ muon decays, with $\sin^2 2\theta_{13} = 0.004$
and the other oscillation parameters specified by Eq.~(\ref{eq:params}).}
\label{table:s_1}
\smallskip
\begin{tabular}{cccccccc}
& sign of& \multicolumn{3}{c}{$E_{\mu^+}$ (GeV)}&
\multicolumn{3}{c}{$E_{\mu^-}$ (GeV)}\\
& $\delta m^2_{32}$& 10& 30& 50& 10& 30& 50\\
\hline
\multicolumn{8}{c}{$L=732$ km}\\
$\nu_\mu$ (no osc)& & 14,300& 382,000& 1,780,000& 29,000& 772,000& 3,560,000\\
$\nu_\mu\to\nu_\mu$& $+$& 11,200& 372,000& 1,760,000& 22,800& 750,000&
3,520,000\\
& $-$& 11,200& 372,000& 1,760,000& 22,900& 751,000& 3,530,000\\
$\nu_e$ (no osc)& & 24,200& 656,000& 3,050,000& 12,400& 329,000& 1,520,000\\
$\nu_e\to\nu_\mu$& $+$& 19.4& 73.6& 127& 8.6& 35.2& 61.8\\
& $-$& 8.4& 30.2& 52.2& 5.34& 16.6& 27.8\\
\hline
\multicolumn{8}{c}{$L=2800$ km}\\
$\nu_\mu$ (no osc)& & 1140& 27,500& 123,000& 2200& 52,500& 244,000\\
$\nu_\mu\to\nu_\mu$& $+$& 168& 18,800& 107,000& 276& 35,900& 212,000\\
&  $-$& 166& 18,900& 107,000& 274& 36,200& 212,000\\
$\nu_e$ (no osc)& & 1400& 43,300& 211,000& 900& 22,600& 102,000\\
$\nu_e\to\nu_\mu$& $+$& 5.2& 56.4& 108& 0.82& 11.6& 29\\
&  $-$& 0.64& 10& 24.6& 4.4& 19.2& 29.6\\
\hline
\multicolumn{8}{c}{$L=7332$ km}\\
$\nu_\mu$ (no osc)& & 152& 3800& 17,500& 294& 7660& 35,600\\
$\nu_\mu\to\nu_\mu$& $+$& 84& 586& 7140& 152& 1100& 14,600\\
& $-$& 84& 586& 7220& 158& 1200& 15,000\\
$\nu_e$ (no osc)& & 218& 6680& 30,500& 120& 3300& 15,200\\
$\nu_e\to\nu_\mu$& $+$& 2.58& 30.2& 32.8& 0.066& 0.89& 1.98\\
& $-$& 0.112& 1.77& 3.8& 1.52& 14.6& 144
\end{tabular}
\end{table}

\begin{table}
\caption[]{Statistical precision in determining
$\sin^2 2\theta_{23}$ and $\delta m^2_{32}$ if
$\sin^2 2\theta_{23}=1.0$ and $\delta m^2_{32}=3.5\times10^{-3}{\rm\,eV}^2$.
Fit results are tabulated
for various muon storage ring energies and baselines. \label{table:limits}}
\begin{tabular}{cccccc}
 & & & Fitted energy&sin$^22\theta_{23}$&$\delta m^2_{32}$ \\
$E_\mu$ (GeV) & $L$ (km) & Events fitted &  range (GeV) &
$\%$ error &   $\%$ error \\ \hline
10 & 732  & 847 & 0--4 &  7.6 &  6.7   \\
10 & 2800 & 284 & 0--10 &  1.1 &  2.4   \\
10 & 7332 & 126 & 0--12 & 13 &  6.3   \\
30 & 732  & 3984 & 0--12 & 14 &  8.9   \\
30 & 2800 & 623 & 0--12  & 2.0 &  3.2   \\
30 & 7332 & 655 & 0--25 &  0.57 &  1.2   \\
50 & 732  & 1573 & 0--12 & 17 &  12   \\
50 & 2800 & 169 & 0--10 & 1.8 &  4.9   \\
50 & 7332 & 834 & 0--28 & 0.64 &  1.4   \\
\end{tabular}
\end{table}

\begin{table}
\caption{Summary of sensitivity versus baseline and
stored muon energy.}
\begin{tabular}{ccccccc}
& & \multicolumn{2}{c}{Survival}& \multicolumn{3}{c}{Appearance}\\
\cline{3-4} \cline{5-7}\\[-2ex]   
& &sin$^22\theta_{23}$&$\delta
m^2_{32}$&sin$^22\theta_{13}$&sin$^22\theta_{13}$& \\
& &statistical&statistical& 10~Event&3$\sigma$ sign& $\delta m_{21}^2$\\
$L$ (km)& $E_\mu$ (GeV)&precision&precision&sensitivity&$\delta m^2$&
effects\\
\hline
732&  10&7.6\% &6.7\%&0.002 &$>$0.1& large\\
732&  30&14\%  &8.9\%&0.0005&0.1   & large\\
732&  50&17\%  &12\%&0.0003 &$>0.1$& large\\
2800& 10&1.1\% &2.4\%&0.008 &0.1   & moderate\\
2800& 30&2.0\% &3.2\%&0.0007&0.005 & moderate\\
2800& 50&1.8\% &4.9\%&0.0004&0.003 & moderate\\
7332& 10&13\%  &6.3\%&0.02  &$>$0.1& negligible\\
7332& 30&0.57\%&1.2\%&0.001 &0.04  & negligible\\
7332& 50&0.64\%&1.4\%&0.002 &0.02  & negligible
\end{tabular}
\end{table}


\newpage

\newpage

\newpage

\newpage

\end{document}